\documentclass[aps,prd,preprint,superscriptaddress]{revtex4-2}
\usepackage{amsmath}
\usepackage{hyperref}
\usepackage{graphicx}
\usepackage{amsfonts}
\usepackage{physics}
\title{Scrambling In Charged Hairy Black Holes and the Kasner Interior}

\date{\today}

\begin{document}

\title{Scrambling in charged hairy black holes and the Kasner interior}
\author{Hadyan Luthfan Prihadi}

\email{hady001@brin.go.id}
\affiliation{Research Center for Quantum Physics, National Research and Innovation Agency (BRIN), South Tangerang 15314, Indonesia.}
\author{Donny Dwiputra}
\email{donny.dwiputra@apctp.org}
\affiliation{Asia Pacific Center for Theoretical Physics, Pohang University of Science and Technology, Pohang 37673, Gyeongsangbuk-do, South Korea.}
\affiliation{Research Center for Quantum Physics, National Research and Innovation Agency (BRIN), South Tangerang 15314, Indonesia.}
\author{Fitria Khairunnisa}
\email{30223301@mahasiswa.itb.ac.id}
\affiliation{Theoretical High Energy Physics Group, Department of Physics, FMIPA, Institut Teknologi Bandung, Jl. Ganesha 10 Bandung, Indonesia.}
\affiliation{Research Center for Quantum Physics, National Research and Innovation Agency (BRIN), South Tangerang 15314, Indonesia.}
\author{Freddy Permana Zen}
\email{fpzen@itb.ac.id}
\affiliation{Theoretical High Energy Physics Group, Department of Physics, FMIPA, Institut Teknologi Bandung, Jl. Ganesha 10 Bandung, Indonesia.}
\affiliation{Indonesia Center for Theoretical and Mathematical Physics (ICTMP), Institut Teknologi Bandung, Jl. Ganesha 10 Bandung,
	40132, Indonesia.}
\begin{abstract}
    We analyze how the axion parameter, the Einstein-Maxwell-Scalar (EMS) coupling constant, and the charge density affect the chaotic properties of a charged hairy black hole, as characterized by the quantum Lyapunov exponent. We inject charged shock waves from the asymptotic boundary and compute the out-of-time-ordered correlators (OTOCs). Due to the relevant deformation in the boundary theory induced by a bulk scalar field, the bulk solution flows to a more general Kasner spacetime near the black hole singularity. We examine the behavior of chaotic parameters, including the Lyapunov exponent, butterfly velocity, and scrambling time delay, under this deformation. We find that as the deformation parameter increases, the ratio of the quantum Lyapunov exponent to the surface gravity decreases. For sufficiently large deformation, the Lyapunov exponent in the deformed geometry can exceed that of the axion Reissner-Nordström case. We observe that boundary deformation generally reduces the scrambling time delay, with the EMS coupling having a significant effect on the delay. These results provide further insight into the role of boundary deformations in modifying chaotic properties in charged hairy black holes.
\end{abstract}
\maketitle
\newpage
\tableofcontents
\section{Introduction}
One of the goals of the AdS/CFT correspondence \cite{Maldacena1997_LargeN} is to reconstruct spacetime geometry in the bulk from the boundary CFT data (see, for example, \cite{deHaro2001,DeJonckheere2017}). Initially, observables at the boundary, such as the expectation value of the boundary field, denoted by $\expval{\mathcal{O}}$, and the energy-momentum tensor, $\expval{T_{tt}}$, can reconstruct the spacetime geometry close to the boundary. It is then noticed that the holographic entanglement entropy calculated by the Ryu-Takayanagi entangling surface \cite{RyuTakayanagi1,RyuTakayanagi2} can penetrate deeper into the bulk region. Furthermore, some entangling surfaces stretch from the left asymptotic boundary to the right one through the entire bulk spacetime and even go into the interior region of a black hole \cite{Hartman2013}. This surface may shed light on understanding the black hole interior from the boundary perspective.\\
\indent The entangling surface stretching between the left and right asymptotic boundaries has been used to study the chaotic behavior of a black hole perturbed by the gravitational shock waves \cite{Shenker2014,Shenker2015stringy,Roberts2015}. This surface is utilized to calculate mutual information, $I(A;B)$, of two subregions, denoted by $A$ and $B$, that live in the left and right boundary CFT, respectively. The result shows that the black hole amid chaotic behavior with the out-of-time-ordered correlator (OTOC) vanishes in a time scale that is logarithmically dependent on the black hole entropy \cite{Sekino2008}. The exponential behavior of the vanishing OTOC is controlled by a parameter called the quantum Lyapunov exponent, $\lambda_L$, it is analogous to the Lyapunov exponent in a classical chaotic system. The Lyapunov exponent of a black hole is equal to the surface gravity of the black hole and hence saturates the Maldacena-Shenker-Stanford chaos bound \cite{Maldacena2016bound}. 
The chaotic behavior of various black holes, including charged and rotating black holes, has been extensively studied using holography in \cite{Leichenauer2014,Jahnke2018,Jahnke2019,Horowitz2022,Malvimat2022,Malvimat2023,Malvimat2023KerrAdS4,Prihadi2023,Prihadi2024}.\\
\indent The AdS black hole spacetime in the bulk may also contain a scalar field producing the solution of a hairy black hole. In the AdS/CFT dictionary, this scalar field plays an important role in generating deformation in the boundary theory. This deformation creates a holographic renormalization group flow from the UV theory on the boundary to the IR theory near the black hole horizon. If we go deeper into the interior of a black hole, i.e., going into the trans-IR region where the energy scale becomes imaginary, we may arrive in a regime where the geometry becomes Kasner spacetime near the black hole singularity \cite{Hartnoll2020,Frenkel2020,sword_kasner_2022,sword_what_2022}. Recent works also include the interior structure of AdS black holes with charged hair \cite{carballo2024divinginsideholographicmetals}, stringy corrections to the Kasner interiors \cite{Caceres2024eons} and the dynamical analysis of black hole interior near the singularity in the Mixmaster model \cite{Oling2024}. The quantities in the Kasner spacetime can, in principle, be determined by the boundary data by solving the equations of motion either analytically or numerically. Recently in \cite{caceres_shock_2024}, the Kasner spacetime in the interior has been studied under the influence of some gravitational shock waves that can probe chaos.\\
\indent In this work, we extend previous studies on black hole chaos by considering the injection of charged shock waves instead of neutral ones. In a charged black hole background, these shock waves can interact with the black hole's charge, introducing new dynamical effects in the scrambling process such as the scrambling time delay \cite{Horowitz2022}. To incorporate this interaction, we add a Maxwell field to the theory, leading to a charged hairy black hole where the bulk scalar field couples to the $U(1)$ gauge field $A_\mu$. This setup has been extensively studied as a holographic model of superconductors \cite{Hartnoll2020,sword_kasner_2022,sword_what_2022}, making it particularly interesting to explore its chaotic properties. Given that holographic superconductors are based on black hole solutions, they may also exhibit chaotic behavior, which motivates us to investigate how their interior structure is influenced by the Lyapunov exponent as a diagnostic of chaos in the boundary theory. Specifically, we analyze the relation between chaotic properties—such as the Lyapunov exponent and scrambling time delay—and the boundary deformation parameter, which modifies the interior geometry into a more general Kasner spacetime. This allows us to gain insight into the connection between boundary chaos and the black hole interior. Additionally, aside from the scrambling time studied earlier in \cite{caceres_shock_2024}, the instantaneous Lyapunov exponent can also be extracted from the mutual information $I(A;B)$, calculated holographically \cite{Malvimat2023KerrAdS4,Malvimat2023,Prihadi2023}.
 \\
\indent The holographic model in \cite{sword_kasner_2022} also contains the Einstein-Maxwell-Scalar (EMS) coupling term and the axion field. The axion scalar terms were introduced in \cite{Andrade2015} in the context of holographic superconductors to break translational invariance in the boundary theory. Furthermore, the axion field is also present in the study of holographic phonon \cite{Alberte2018phonon} which generates finite graviton mass and plays a role in the spontaneous breaking of the translational symmetry as well. On the other hand, the EMS term was introduced in \cite{Dias2021JHEP} to study the Kasner interior of an asymptotically flat black hole. We study how these parameters: axion parameter, EMS coupling constant, and charge density, influence the chaotic behavior of the black hole in the interior, since these parameters highly influence the interior structure of the black hole \cite{Mansoori2021,Dias2021JHEP,Arean2024Kasner}. \\
\indent Other than the quantum Lyapunov exponent, we also consider the butterfly velocity of the black hole when it is perturbed by the localized shock waves \cite{Roberts2015,Jahnke2018,Prihadi2024}. We study how this velocity gets affected by the relevant deformation in the boundary and investigate its relation with the Kasner exponent. Although the butterfly velocity only depends on the radius at the black hole horizon, this chaotic quantity is still related to the OTOC. We expect to find a non-trivial relation between butterfly velocity and both the boundary deformation and Kasner exponent and see how the parameters $\zeta,\gamma,\rho$ affect this relation.\\
\indent Once we inject charged shock waves instead of neutral ones into the black hole, we then expect the shock waves to bounce in the interior region due to the interaction between the black hole charge and the shock wave charge. This bouncing phenomenon was initially studied in a Reissner-Nordstr\"om-AdS black hole background \cite{Horowitz2022} and later extended to charged rotating black hole in the Einstein-Maxwell dilaton-axion theory \cite{Prihadi2023}. Since the bounce happens inside the horizon, it is important to study its relevance with the emergence of Kasner spacetime. The parameters $\zeta,\gamma,\rho$ might also influence the scrambling time delay, especially for $\gamma$ since it mainly controls the coupling between the scalar field and the electromagnetic potential.\\
\indent The structure of this paper is as follows. In Section 2, we briefly review Kasner geometry in the charged hairy black hole model with EMS coupling and axion field in \cite{sword_kasner_2022}. We also show the relation between the Kasner exponent $p_t$ and the boundary deformation $\phi_0/T$ under the variation of $\zeta,\gamma,\rho$. In Section 3, we calculate mutual information holographically as it is perturbed by charged gravitational shock waves. We extract the instantaneous Lyapunov exponent and study how it is influenced by the presence of boundary deformation. We then study the influence of the $\zeta,\gamma,\rho$ parameters to the Lyapunov exponent as it is deformed by the boundary deformation. The relationship between the Lyapunov and Kasner exponents, as well as how these parameters influence them, are also adressed. Furthermore, we study the impacts of the given parameters on butterfly velocity when the black hole is perturbed by localized shock waves. In Section 4, we analyze how the scrambling time delay gets affected by the boundary deformation parameter, and the parameters $\zeta,\gamma,\rho$. We also see how the coupling parameter $q$ between the scalar field and the gauge field affect Finally, we summarize our results and presents some discussions in Section 5.
\section{Interior of Charged Hairy Black Hole}
We use the charged hairy black hole model containing the massive Klein-Gordon field $\phi$, the Maxwell field $A_\mu$, and the massless axion field $\Omega$, along with the Einstein-Maxwell-scalar coupling term, which directly couples $F_{\mu\nu}F^{\mu\nu}$, where $F_{\mu\nu}=\partial_\mu A_\nu-\partial_\nu A_\mu$, with the scalar field $\phi$. This model was previously studied by \cite{sword_kasner_2022} to investigate Kasner spacetime in the interior of holographic superconductors. The total action of the four-dimensional model is given by $S=S_1+S_2+S_3$, with action:\\
\begin{align}
    S_1&=\int d^4x\sqrt{|g|}\qty(R+\frac{6}{L^2}),\\
    S_2&=\int d^4x\sqrt{|g|}\bigg(-\frac{L^2}{4}F_{\mu\nu}F^{\mu\nu}-g^{\mu\nu}\qty(\partial_\mu\phi-iqA_\mu\phi)(\partial_\nu\phi^*+iqA_\nu\phi^*)-\frac{1}{2}m^2|\phi|^2\bigg),\\
    S_3&=\int d^4x\sqrt{|g|}\qty(-\frac{L^2}{4}\gamma F_{\mu\nu}F^{\mu\nu}|\phi|^2-\frac{1}{L^2}K(X)),
\end{align}
where $L$ is the AdS radius, $q$ is the constant of coupling between $A_\mu$ and $\phi$, and $\gamma$ is the EMS coupling constant. We also introduce the K-essence $K(X)$ with
\begin{equation}
    X=\frac{L^2}{2}=\sum_I g^{\mu\nu}\partial_\mu\Omega^I\partial_\nu\Omega^I.
\end{equation}
Following \cite{sword_kasner_2022}, the axion field is defined as
\begin{equation}
    \Omega^I=\zeta x^I,
\end{equation}
with $I=(x,y)$ for some axion parameter $\zeta$. In this case, $X=\zeta^2r^2$ and the polynomial form of the K-essence is then given by 
\begin{equation}
    K(X)=X^n=(\zeta^2r^2)^n.
\end{equation}
In the numerical calculations, we mainly focus on $n=1$ case.\\
\indent We work with a gauge in which the scalar field is real and we assume that it only depends on $r$, 
\begin{equation}
    \phi=\phi^*=\phi(r).
\end{equation}
We also choose the Maxwell field to describe only radially dependent electric potential,
\begin{equation}
    A_\mu dx^\mu=A_t dt=\Phi(r) dt.
\end{equation}
Both $\phi(r)$ and $\Phi(r)$ create the charged, asymptotically AdS hairy black hole with metric given by the following ansatz,
\begin{equation}\label{metricansatz}
    ds^2=\frac{L^2}{r^2}\qty(-f(r)e^{-\chi(r)}dt^2+\frac{dr^2}{f(r)}+dx^2+dy^2).
\end{equation}
In this metric, $r$ is the AdS radial coordinate, where $r\rightarrow0$ corresponds to the AdS boundary, where the CFT resides, and $r\rightarrow\infty$ signifies the black hole singularity in the interior.\\
\indent The equations of motion now consist of the Klein-Gordon equation, the Maxwell equation, and the $tt$ and $rr$ components of the Einstein equations which are respectively given by\\
\begin{align}
        \phi''+\phi'\qty(-\frac{2}{r}+\frac{f'}{f}-\frac{\chi'}{2})+\phi\qty(-\frac{L^2m^2}{r^2f}+\frac{q^2 e^{\chi}\Phi^2}{f^2}+\frac{\gamma r^2 e^\chi \Phi'^2}{2f})&=0,\\
    \Phi''+\Phi'\frac{\qty((\gamma\phi^2+1)\chi'+4\phi\phi')}{2(1+\gamma\phi^2)}-\Phi\frac{2q^2\phi^2}{r^2f(1+\gamma\phi^2)}&=0,\\
    \frac{q^2e^\chi\Phi^2\phi^2}{f^2}-\frac{\chi'}{r}+\phi'^2&=0,\\
    \frac{6}{r^2}-\frac{6}{r^2f}-\frac{2f'}{rf}+\frac{L^2m^2\phi^2}{r^2f}+\frac{K(X)}{r^2f}+\frac{q^2e^\chi\phi^2\Phi^2}{f^2}+\frac{r^2e^\chi(\gamma\phi^2+1)}{2f}(\Phi')^2+\phi'^2&=0.
\end{align}
There are two roots of $f(r)=0$, which gives us the black hole horizon radius $r_h$ and the axion-Reissner-Nodrstr\"om horizon $r_{\text{aRN}}$. The $r_{\text{aRN}}$ serves as the inner horizon of the black hole. However, as first pointed out in \cite{Hartnoll:2020rwq}, the scalar field in the bulk generates a black hole without an inner horizon due to the collapse of the Einstein-Rosen bridge. This collapse is also found in the model studied in \cite{sword_kasner_2022}.
\subsection{Boundary Conditions from the CFT}
In this subsection, we solve the equations of motion for $\phi,\Phi,f,\chi$ numerically. The equations of motion are second order in both $\phi(r)$ and $\Phi(r)$ and hence we need two boundary conditions for those fields. However, they are first order in $f(r)$ and $\chi(r)$ and thus only one boundary condition for each field is needed. We integrate the equation of motions numerically from $r=r_h-\delta$ to the boundary at $r=0$ and from $r=r_h+\delta$ to the interior at $r\gg 1$, where $\delta$ is a small number. Therefore, we need to specify proper initial conditions at the horizon for all fields. Since $f(r)$ is the blackening factor for the black hole solution, we have $f(r_h)=0$. The electric potential $\Phi(r)$ needs to vanish at the horizon as well so that the norm of $A_\mu$ is finite there.\\
\indent Other horizon quantities can be determined by expanding the fields near $r_h$,
\begin{align}
    \phi&=\phi_{h1}+\phi_{h2}(r-r_+)+\phi_{h3}(r-r_+)^2+...\;,
    \end{align}
    \begin{align}
    f&=f_{h1}(r-r_h)+f_{h2}(r-r_h)^2+...\;,
    \end{align}
    \begin{align}
    \chi&=\chi_{h1}+\chi_{h2}(r-r_h)+\chi_{h3}(r-r_h)^2+...\;,
    \end{align}
    \begin{align}
    \Phi&=\Phi_{h1}(r-r_h)+\Phi_{h2}(r-r_h)^2+\Phi_{h3}(r-r_h)^3+...\;.
\end{align}
Plugging these expressions to the equations of motion and set the coefficient of the divergent ($\mathcal{O}(1/(r-r_h)$) parts to zero, we will see that only three quantities are independent. In the numerical calculations, we choose $\phi_{h1}, \Phi_{h1},\chi_{h1}$ to be the independent quantities for the boundary conditions at the horizon, while the other parameters can be expressed in terms of $\phi_{h1}, \Phi_{h1},\chi_{h1}$.\\
\indent Although we need to specify the boundary condition at the horizon for numerical purposes, we need to fix the value of the fields in the asymptotic boundary, where the CFT lives. This is achieved using the shooting method. The solutions of the field equations near the boundary $r\rightarrow0$ are given by
\begin{align}
    \phi&=\phi_{(1)}r+\phi_{(2)}r^2+...\;,\\
    f&=1+f_{(1)}r^2+f_{(2)}r^3+...+\tilde{f}r^{2n}+...\;,\\
    \chi&=\chi_{(1)}+\frac{1}{2}\phi_{(2)}^2r^2+\frac{4}{3}\phi_{(1)}\phi_{(2)}r^3+...\;,\\
    \Phi&=\Phi_{(1)}+\Phi_{(2)}r+q^2\phi_{(1)}^2\Phi_{(1)}r^2+...\;.
\end{align}
These expansion coefficients determine the boundary data from the CFT theory. For example, in the standard quantization scheme, $\phi_{(1)}\equiv\phi_0$ is the boundary deformation which plays the role as a source to the boundary scalar field $\mathcal{O}$ that lives in the dual CFT theory. On the other hand, $\phi_{(2)}$ is the expectation value $\expval{\mathcal{O}}$ as the response of this deformation. Note that we choose $m^2=-2/L^2$, which satisfies the Breitenlohner-Freedman bound and generates relevant deformation in the CFT. For the Maxwell field, $\Phi_{(1)}\equiv\mu$ determines the chemical potential while $\Phi_{(2)}\equiv-\rho$ determines the charge density of the theory. This charge density will play an important role in making the black hole solution a charged hairy black hole, which reduces to the axion-Reissner-Nordstr\"om (aRN) solution when the scalar field $\phi$ is absent.
\\
\indent Since the boundary data $\phi_{(1)}$ and $\Phi_{(2)}$ are fixed, we use shooting method to find suitable values of $\phi_{h1}$ and $\Phi_{h1}$ that gives us correct results at the boundary, i.e., we find the correct value $\phi_{h1}$ and $\Phi_{h1}$ that gives us
\begin{equation}
    \lim_{r\rightarrow 0}\frac{\phi(r)}{r}=\phi_0,
\end{equation}
and
\begin{equation}
    \lim_{r\rightarrow0}\Phi'(r)=-\rho.
\end{equation}
Another boundary data that is also important is $\chi_{(1)}$. For an asymptotically AdS spacetime, we need this value to approach zero in the boundary. To achieve this, we can choose $\chi_{h1}$ arbitrarily and rescale the fields as
\begin{equation}
    e^\chi\rightarrow a_1^2e^\chi,\;\;\;t\rightarrow a_1 t,\;\;\;\Phi\rightarrow\frac{\Phi}{a_1},
\end{equation}
so that we have $\chi(0)=0$, by choosing $a_1=e^{-\chi(0)/2}$. This scaling does not change the equations of motion and we call this the time scaling symmetry, which sets the black hole temperature to
\begin{equation}
    T_{BH}=\frac{|f'(r_h)|e^{-(\chi(r_h)-\chi(0))/2}}{4\pi}.
\end{equation}
\subsection{Flow Through General Kasner Interior}
\begin{figure}
    \centering
    \includegraphics[width=0.32\linewidth]{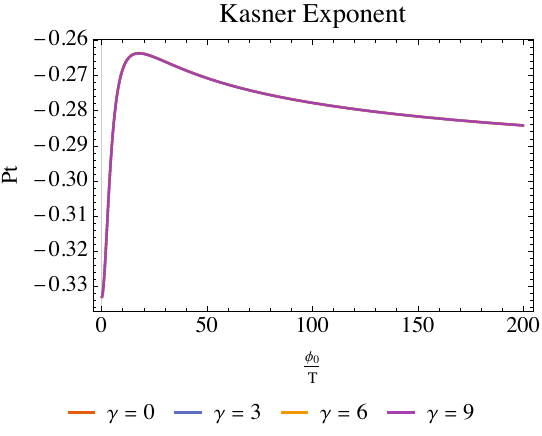}
    \includegraphics[width=0.32\linewidth]{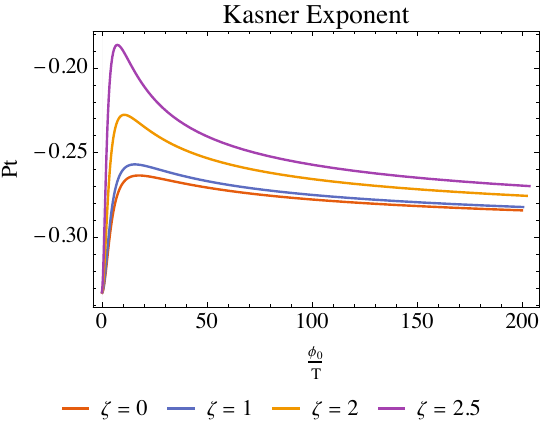}
    \includegraphics[width=0.32\linewidth]{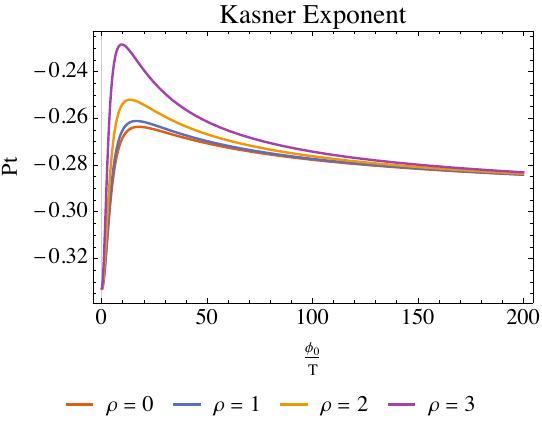}
    \caption{The plot for the Kasner exponent $p_t$ versus boundary deformation $\phi_0/T$ when we vary $\gamma$ (left), $\zeta$ (center), and $\rho$ (right). In this plot, we use $q=0.1$.}
    \label{fig:plotkasner}
\end{figure}

\begin{figure}
    \centering
    \includegraphics[width=0.32\linewidth]{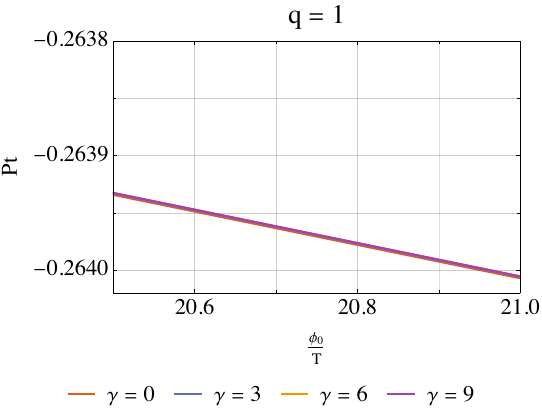}
    \includegraphics[width=0.32\linewidth]{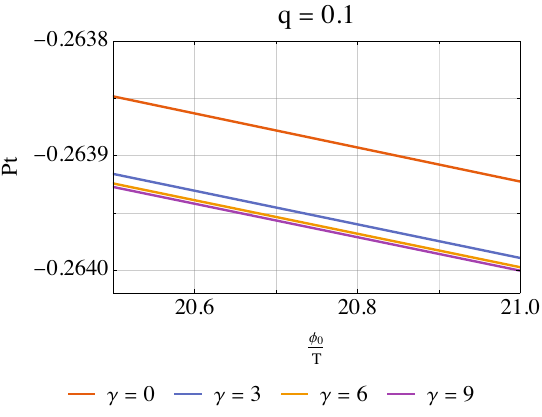}
    \includegraphics[width=0.32\linewidth]{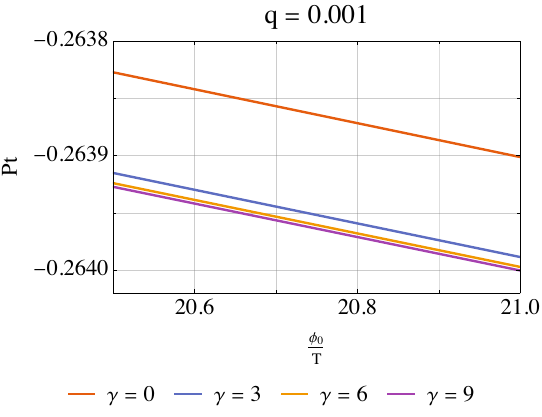}
    \caption{The plot for the Kasner exponent $p_t$ versus boundary deformation $\phi_0/T$ when we vary $\gamma$. We use $q = 1$ (left), $q=0.1$ (center), and $q=0.001$ (right).}
    \label{fig:kasnergammazoom}
\end{figure}

The scalar field $\phi(r)$ propagates through the AdS bulk spacetime and generates relevant deformation $\phi_0$ in the boundary CFT theory such that the action of the boundary theory is deformed as
\begin{equation}
    \delta S=\int d^3x\phi_0\mathcal{O},
\end{equation}
where $\mathcal{O}$ is the scalar field in the dual theory. This deformation generates a holographic RG flow from the UV theory in the boundary through the IR theory in the bulk (near the horizon). We can also penetrates through the interior of the black hole, where the theory becomes trans-IR \cite{caceres2022trans}. In this regime, the energy scale becomes imaginary and the radial coordinate becomes timelike.\\
\indent As we go through the interior and approach the singularity, where $r\gg1$, the fields behave as
\begin{align}\label{nearsingularityexpansion1}
    \phi&=\sqrt{2}c\log r+...\;,\\
    f&=-f_{K1}r^{3+c^2}+...\;,\\
    \chi&=2c^2\log r+\chi_{K1}+...\;,\\
    \Phi&=\Phi_{K1}r^{1-c^2}+\Phi_{K2}+...\;,\label{nearsingularityexpansion4}
\end{align}
when $\gamma=\zeta=0$. With these expansions, and the radial coordinate reparameterization $r=\tau^{-2/(3+c^2)}$, the metric in the deep interior becomes Kasner spacetime, which can be written as
\begin{equation}
    ds^2=-d\tau^2+a_t\tau^{2p_t}dt^2+a_x\tau^{2p_x}(dx^2+dy^2),
\end{equation}
and
\begin{equation}
    \phi\sim-p_\phi\log\tau+\phi_\tau,
\end{equation}
where $a_t,a_x,\phi_\tau$ are constants. The exponents $p_t$ and $p_x$ are the Kasner exponents, which along with $p_\phi$, are given by
\begin{equation}
    p_x=\frac{2}{3+c^2},\;\;\;p_t=\frac{c^2-1}{3+c^2},\;\;\;p_\phi=\frac{2\sqrt{2}c}{3+c^2}.
\end{equation}
\indent All of the Kasner exponents only depends on the integration constant $c$ near the singularity. This integration constant can be obtained from the boundary conditions in the UV boundary $r\rightarrow0$ using numerical calculations. After we obtain the fields numerically, $c$ can be extracted from $\frac{r}{\sqrt{2}}\frac{d\phi}{dr}$ at large enough $r$. Therefore, if we find that $rd\phi/dr$ approaches constant for some large value of $r$, we can say that, in this region, we approach the Kasner regime. We plot how one of the Kasner exponents $p_t$ as a function of dimensionless boundary deformation $\phi_0/T$ after varying $\gamma,\zeta,$ and $\rho$ in Figure \ref{fig:plotkasner}.\\
\indent The effect of varying $\gamma$ on the Kasner exponent is very subtle for $q = 0.1$. If we rescale the scalar field and the electromagnetic field as $\phi \to \phi / q$ and $\Phi \to \Phi / q$, the action of the model transforms into
\begin{align}
    S=&\int d^4x\sqrt{|g|}\bigg\{\bigg(R+\frac{6}{L^2}-\frac{1}{L^2}K(X)\bigg)\\\nonumber
    &+\frac{1}{q^2}\bigg(-\frac{L^2}{4}F_{\mu\nu}F^{\mu\nu}-g^{\mu\nu}\qty(\partial_\mu\phi-iqA_\mu\phi)\\\nonumber
    &\times(\partial_\nu\phi^*+iqA_\nu\phi^*)-\frac{1}{2}m^2|\phi|^2\bigg)\\\nonumber
    &+\frac{1}{q^4}\bigg(-\frac{L^2}{4}\gamma F_{\mu\nu}F^{\mu\nu}|\phi|^2\bigg)\bigg\}.
\end{align}
 In the probe limit (when $q\gg 1$), the contribution of $\gamma$ to the Kasner exponent is suppressed compared to the leading terms in the action, making its effect negligible. The terms containing $\gamma$ in the equation of motions under this scaling when $q$ is large behave as $\mathcal{O}(1/q^4)$. However, as shown in Figure \ref{fig:kasnergammazoom}, for smaller $q$, where the backreaction of the gauge and scalar fields is more significant, the influence of $\gamma$ on the Kasner exponent becomes more apparent although it is still weak. This is because the coupling $\gamma$ primarily governs how the scalar field, through $\phi^2$, affects the Maxwell field $\Phi^2$, while the Kasner exponent is more determined by the bulk solution of $\phi$. Consequently, this also impacts how $\gamma$ controls other chaotic parameters, which are mainly influenced by the near-horizon metric functions such as $f(r)$ and $\chi(r)$, as will be discussed in the next section.
\section{Mutual Information and Lyapunov Exponent}
\begin{figure}
    \centering
    \includegraphics[scale=0.5]{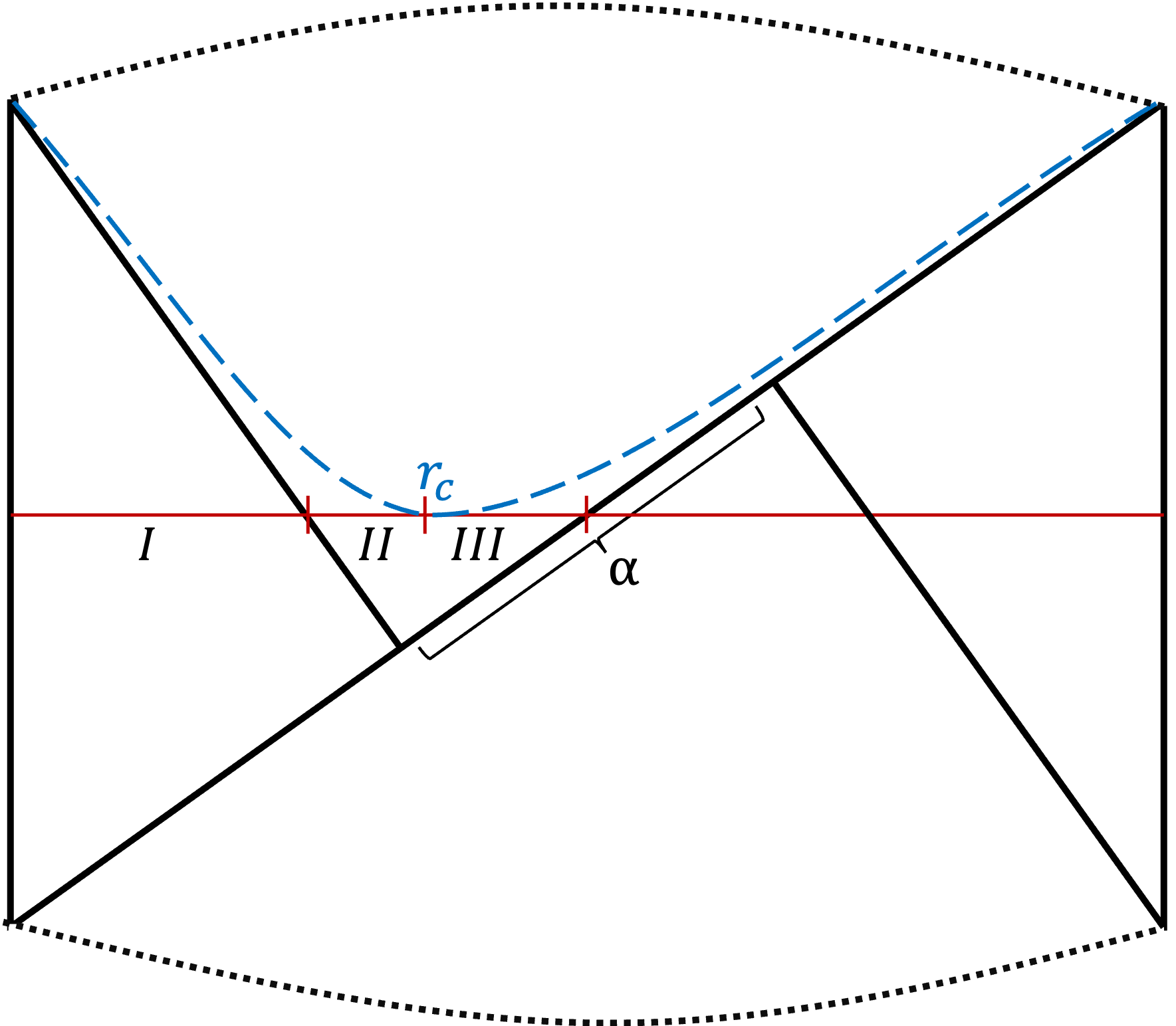}
    \caption{Penrose diagram of the black hole spacetime perturbed by gravitational shock waves $\alpha$. $r_c$ represents the critical turning point while the horizontal red line represents the minimal surface. $I,II,$ and $III$ are the segments that we are using in calculating the minimal surface.}
    \label{fig:HEEpicture}
\end{figure}

In this section, we investigate the chaotic behavior of charged hairy black hole using holographic calculations as first done in \cite{Shenker2014,Leichenauer2014}. We calculate the OTOC in the scrambling regime by calculating the mutual information $I(A;B)=S_A+S_B-S_{A\cup B}$ using holographic entanglement entropy \cite{RyuTakayanagi1,RyuTakayanagi2,hubeny2007,Hartman2013}. The calculation of the entanglement entropy $S_{A\cup B}$ involves the calculation of the entangling RT/HRT surfaces which stretch from left to right asymptotic boundaries \cite{Hartman2013}. This surface penetrates through the interior of the black hole. We calculate the area of this surface in a background that is perturbed by charged gravitational shock waves \cite{Horowitz2022,Prihadi2023} with shock wave parameter given by
\begin{equation}
    \alpha=\frac{\beta E_0}{S}(1-\mu\mathcal{Q})e^{\frac{2\pi t_w}{\beta}},
\end{equation}
where $\mu$ is the black hole's electric potential and $\mathcal{Q}$ is the shockwaves' charge per unit energy. The electric potential (or chemical potential) $\mu$ can be obtained from the boundary limit $\lim_{r\rightarrow0}\Phi(r)$. Here, $t_w$ is the time when the perturbation with energy $E_0$ is sent from the left boundary. The shockwave parameter $\alpha$ shifts the background metric as shown in Figure \ref{fig:HEEpicture}.\\
\indent The mutual information $I(A;B)$ is bounded below by the correlation function $\expval{\mathcal{O}_A\mathcal{O}_B}_w$ \cite{Wolf2008}, where the state is perturbed at $t_w$. The operators $\mathcal{O}_A$ and $\mathcal{O}_B$ are local operators that act in the subregion $A$ in the left boundary and the subregion $B$ in the right boundary, respectively. The CFT boundary state that is considered here is the entangled thermofield double state $|TFD\rangle$ that represents charged hairy black holes in the AdS dual. We expect that the mutual information of this TFD state will approach zero exponentially in the scrambling regime. This indicates the chaotic behavior of the charged hairy black hole.\\
\indent It is interesting to study the behavior of the black hole interior from some boundary data that can be determined by $I(A;B)$. This is because the entangling surface of $S_{A\cup B}$ slightly penetrates through the horizon at late times, as we can see later on. Some quantities that can be extracted from this entangling surface is the scrambling time, i.e. the time when $I(A;B)\rightarrow0$, and the Lyapunov exponent $\lambda_L$ that characterizes the chaotic behavior of the black hole. In this work, we focus on the calculation of the Lyapunov exponent, butterfly velocity, and the scrambling time delay. We expect that $\lambda_L$ can give us more insight into studying the interior of this charged hairy black hole by studying the relation between $\lambda_L$ and the boundary deformation $\phi_0/T$ or the Kasner exponent $p_t$. The Lyapunov exponent plays a role in the boundary theory while the Kasner exponent $p_t$ determines the interior solution.
\subsection{Holographic Calculation of Mutual Information}
In this section, we aim to calculate the holographic entanglement entropy $S_A,S_B,$ and $S_{A\cup B}$. We define $A$ and $B$ as two identical subregions in the left and right boundary theories respectively. The subregions $A$ or $B$ can be defined as a region with size $l_x$ in the $x$-direction from $x=0$ to $x=l_x$ and stretches throughout the $y$ direction. The entropy $S_A$ and $S_B$ can be calculated and they are functions of $l_x$. Since $S_A$ and $S_B$ do not go through the horizon, they do not depend on the shock wave parameter $\alpha$.\\
\indent The holographic entanglement entropy $S_{A\cup B}$ is computed by calculating the minimal surface that stretches from the subregion $A$ in the left boundary to the subregion $B$ in the right boundary. This surface penetrates the horizon and thus controlled by the shock wave parameter $\alpha$. Using the metric ansatz in eq. \eqref{metricansatz} for charged hairy black hole, this minimal surface can be calculated by minimizing the area functional
\begin{equation}\label{areafuncitonal}
    \mathcal{A}=L_y\int_0^{r_t} \frac{dr}{r^2}\qty(-fe^{-\chi}t'^2+\frac{1}{f})^{1/2}.
\end{equation}
Here, $L_y$ represents the total length of the $y$-direction, while $r_t$ is the turning point in the bulk where $\dot{r} = 0$. The dot $\dot{}$ indicates differentiation with respect to $t$, whereas the prime $'$ denotes differentiation with respect to $r$. The entanglement entropy $S_{A\cup B}$ is calculated by the sum of the surface areas located at $x=0$ and $x=l_x$.
\\
\indent This area functional is minimized so that the integrand satisfies the Euler-Lagrange equation. Since the metric does not depends explicitly on time, there is a conserved quantity $K$ that satisfy
\begin{equation}\label{Kconstant}
    K=\frac{-fe^{-\chi}}{r^2(-fe^{-\chi}+\dot{r}^2/f^2)^{1/2}}=\frac{(-f(r_t)e^{-\chi(r_t)})^{1/2}}{r_t^2}.
\end{equation}
The constant $K$ is chosen at the turning point $r_t$ where $\dot{r}=0$. The relation between boundary time coordinate and the turning point $r_t$ can be obtained by integrating the time coordinate in Eq. \eqref{Kconstant} from the boundary to the turning point,
\begin{align}\label{timevsr}
t'(r)&=\frac{e^{\chi/2}}{r^2\sqrt{K^{-2}fe^{-\chi}/r^4+1}},\\
    t(r)&=t_b+\int_0^r\frac{e^{\chi/2}dr}{r^2}\frac{1}{\sqrt{K^{-2}fe^{-\chi}/r^4+1}}.
\end{align}
By substituting the expression of $t'$ in \eqref{timevsr} into the area functional in \eqref{areafuncitonal}, we get
\begin{equation}\label{areaminimized}
    \mathcal{A}=L_y\int_0^{r_t}\frac{dr}{r^2\sqrt{f}}\frac{1}{\sqrt{K^2r^4/fe^{-\chi}+1}}.
\end{equation}
We integrate this area from the boundary located at $r=0$ to the turning point inside the horizon $r_t$. The total area that gives the entropy $S_{A\cup B}$ is four times this area.\\
\indent When integrating the area functional, we seperate the integration into three segments using Kruskal coordinates
\begin{align}
    U=e^{\frac{2\pi}{\beta}(r_*(r)-t(r))},\;\;\;V=-e^{\frac{2\pi}{\beta}(r_*(r)+t(r))},
\end{align}
where
\begin{equation}
    r_*=\int_r^0\frac{e^{\chi/2}dr}{f(r)},
\end{equation}
is the tortoise coordinate, following \cite{Leichenauer2014,Malvimat2023KerrAdS4,Prihadi2023}. The first segment is an integration from the boundary with $(U,V)=(1,-1)$ to the horizon with $(U,V)=(U_1,0)$. The second segment stretches from the horizon with $(U,V)=(U_1,0)$ to the turning point at $r=r_t$ denoted by $(U,V)=(U_2,V_2)$. The third segment now runs from the turning point at $(U,V)=(U_2,V_2)$ to the location deformed by the shock waves $\alpha$ at the horizon with $(U,V)=(0,\alpha/2)$. These three segments are depicted in Figure \ref{fig:HEEpicture}.\\
\indent The first segment gives us
\begin{equation}
    U_1^2=\exp\qty[\frac{4\pi}{\beta}\int_0^{r_h}\frac{e^{\chi/2}dr}{f}\qty(1-\frac{1}{\sqrt{K^{-2}fe^{-\chi}/r^4+1}})],
\end{equation}
the second segment gives us
\begin{equation}
    U_2^2=\exp\qty[\frac{4\pi}{\beta}\int_0^{r_t}\frac{e^{\chi/2} dr}{f}\qty(1-\frac{1}{\sqrt{K^{-2}fe^{-\chi}/r^4+1}})],
\end{equation}
and
\begin{equation}\label{persV2}
    V_2=\frac{1}{U_2}\exp\qty[\frac{4\pi}{\beta}\int_{\bar{r}}^{r_t}\frac{e^{\chi/2}dr}{f}],
\end{equation}
while the final segment provides us with the relation between $\alpha$ and the preceding integrals,
\begin{align}\label{alphasegment}
    \frac{\alpha^2}{4V_2^2}&=\exp\bigg[\frac{4\pi}{\beta}\int_{r_t}^{r_h}\frac{e^{\chi/2}dr}{f}\bigg(1-\frac{1}{\sqrt{K^{-2}fe^{-\chi}/r^4+1}}\bigg)\bigg]\\\nonumber
    &=\frac{U_1^2}{U_2^2}.
\end{align}
The radial location $\bar{r}$ that appears in eq. \eqref{persV2} is defined such that the value of $r_*$ at this point is zero inside the horizon. The final integral in eq. \eqref{alphasegment} gives us the the shock wave parameter
\begin{equation}
    \alpha=2\exp\qty(Q_1+Q_2+Q_3),
\end{equation}
where
\begin{align}
    Q_1&=\frac{4\pi}{\beta}\int_{\bar{r}}^{r_t}\frac{e^{\chi/2}dr}{f},\\
    Q_2&=-\frac{2\pi}{\beta}\int_0^{r_h}\frac{e^{\chi/2}dr}{f}\qty(1-\frac{1}{\sqrt{K^{-2}fe^{-\chi}/r^4+1}}),\\
    Q_3&=-\frac{4\pi}{\beta}\int_{r_h}^{r_t}\frac{e^{\chi/2}dr}{-f}\qty(\frac{1}{\sqrt{K^{-2}fe^{-\chi}/r^4+1}}-1).
\end{align}
In contrast with the value of the shock wave parameters found in previous works without scalar hair \cite{Leichenauer2014,Malvimat2022,Malvimat2023KerrAdS4,Prihadi2023}, we find that in our case, $\alpha$ depends on the functions $\chi$ and $f$ and thus it is also controlled by the boundary deformation $\phi_0$. Furthermore, the functions $f(r)$ and $\chi(r)$ are also controlled by other parameters in this charged hairy black hole such as the gauge field coupling constant $q$, axion field strength $\zeta$, and the EMS coupling parameter $\gamma$.\\
\indent We are interested in the limit where $\alpha\rightarrow\infty$. This limit can be achieved by setting the boundary time such that the turning point approaches some value $r_t\rightarrow r_c$, where $r_c$ satisfies
\begin{equation}
    \frac{d}{dr}\qty(\frac{fe^{-\chi}}{r^4})\bigg|_{r=r_c}=0.
\end{equation}
In this limit, both $Q_1$ and $Q_2$ integrals remain finite while $Q_3$ logarithmically diverge as $r_t\rightarrow r_c$. The value of the critical radius $r_c$ can be obtained numerically. As $r_t\rightarrow r_c$, the main contribution of the area functional in eq. \eqref{areaminimized} comes from the region near $r_c$. Therefore, the area is linearly proportional to $Q_3$, and hence depends on $\alpha$ logarithmically,
\begin{align}
    \mathcal{A}&\approx L_y\bigg(\frac{\beta}{4\pi}\bigg)\bigg[\frac{-f(r_c)e^{-\chi(r_c)}}{r_c^4}\bigg]^{1/2}Q_3\\\nonumber
    &\approx L_y\bigg(\frac{\beta}{4\pi}\bigg)\bigg[\frac{-f(r_c)e^{-\chi(r_c)}}{r_c^4}\bigg]^{1/2}\log\alpha.
\end{align}
From the result of the gravitational shock waves parameter $\alpha$, one can see that this area functional grows linearly in the insertion time $t_w$.\\
\indent The entanglement entropy $S_{A\cup B}$ is proportional to the area $\mathcal{A}_{A\cup B}$ that is given by four times the area of a minimal surface $\mathcal{A}$ divided by $4G_N$. Therefore, the mutual information $I(A;B)$ is given by
\begin{align}\label{mutualinformation}
    I(A;B)&=S_A+S_B-S_{A\cup B}\\
    &=\frac{\mathcal{A}_A+\mathcal{A}_B}{4G_N}-\frac{L_y}{2G_N}\qty[\frac{-f(r_c)e^{-\chi(r_c)}}{r_c^4}]^{1/2}t_w\\\nonumber
    &\;\;\;\;+\frac{L_y}{G_N}\qty(\frac{\beta}{4\pi})\qty[\frac{-f(r_c)e^{-\chi(r_c)}}{r_c^4}]^{1/2}\log S\\\nonumber
    &\;\;\;\;+\frac{L_y}{G_N}\qty(\frac{\beta}{4\pi})\qty[\frac{-f(r_c)e^{-\chi(r_c)}}{r_c^4}]^{1/2}\log\frac{1}{1-\mu\mathcal{Q}},
\end{align}
where $\mathcal{A}_A$ and $\mathcal{A}_B$ are the Ryu-Takayanagi surfaces correspond to the subregions $A$ and $B$. The areas $\mathcal{A}_A$ and $\mathcal{A}_B$ does not get influenced by the gravitational shock wave. In calculating this mutual information, we take the energy of the shock waves $E$ to be at the order of a few Hawking quanta so that $\beta E\sim 1$. The mutual information is real-valued since $f(r_c)$ is negative when the critical radius $r_c$ lies inside the horizon. Furthermore, it is also positive-valued since when $\mathcal{A}_{A\cup B}$ becomes greater than $\mathcal{A}_A+\mathcal{A}_B$, the minimal surface $\mathcal{A}_{A\cup B}$ is equal to $\mathcal{A}_A+\mathcal{A}_B$, giving $I(A;B)=0$.\\
\indent The insertion time $t_w$ is in the order of the scrambling time $t_*$ when the mutual information $I(A;B)$ vanishes. This time scale grows logarithmically in the black hole entropy $S$, indicating fast scrambling \cite{Sekino2008}. Other terms such as the $\mathcal{A}_A+\mathcal{A}_B$ and $\log\frac{1}{1-\mu\mathcal{Q}}$ terms do not grow with the entropy of the black hole. However, the latest term contributes to the delay of the scrambling process \cite{Horowitz2022,Prihadi2023} due to bouncing of the shock waves in the interior. 
\subsection{Lyapunov Exponent from Boundary Deformations}
\begin{figure}
    \centering
    \includegraphics[width=0.45\linewidth]{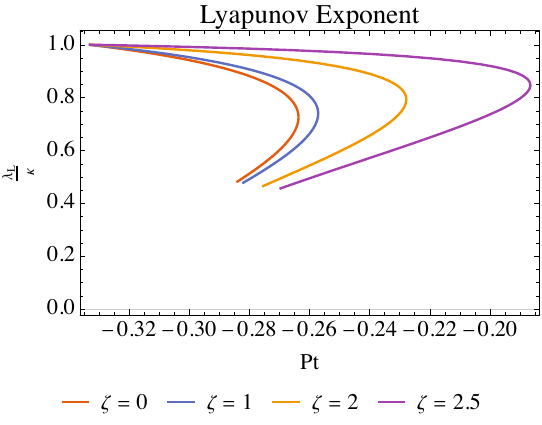}
    \includegraphics[width=0.45\linewidth]{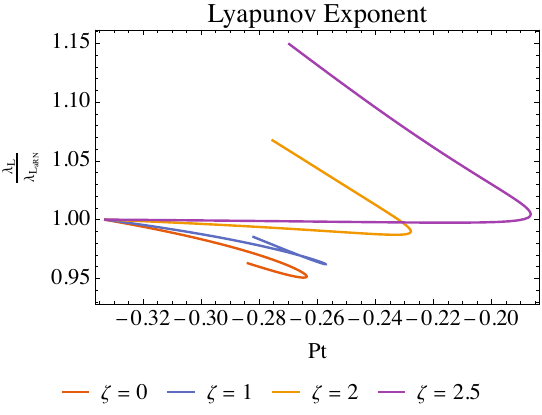}
    \caption{Plots of $\lambda_L/\kappa$ vs $p_t$ (left) and $\lambda_L/\lambda_{L_\text{aRN}}$ vs $p_t$ (right) for various value of the axion field parameter $\zeta$. In this plot, we set $\gamma=0, \rho=0.2,q=0.1$.}
    \label{fig:LSLKKasnerZeta}
\end{figure}

\begin{figure}
    \centering
    \includegraphics[width=0.45\linewidth]{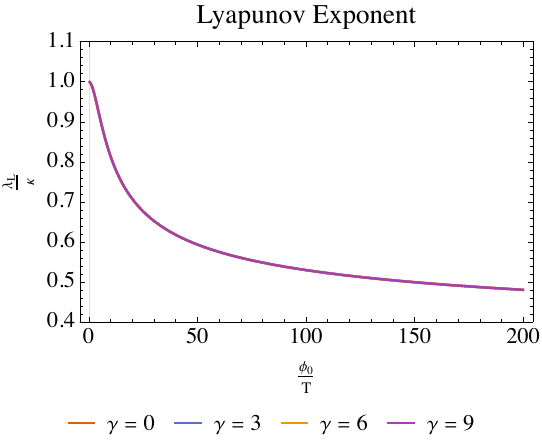}
    \includegraphics[width=0.45\linewidth]{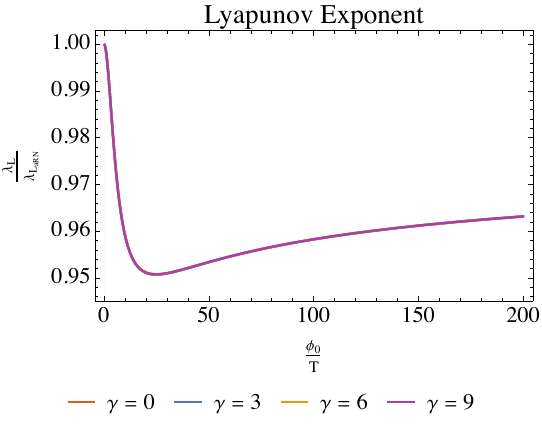}
    \caption{Plots of $\lambda_L/\kappa$ vs $\phi_0/T$ (left) and $\lambda_L/\lambda_{L_\text{aRN}}$ vs $\phi_0/T$ (right) by varying the EMS parameter $\gamma$. In this plot, we set $\zeta=0, \rho=0.2,q=0.1$.}
    \label{fig:LSLKgamma}
\end{figure}

\begin{figure}
    \centering
    \includegraphics[width=0.45\linewidth]{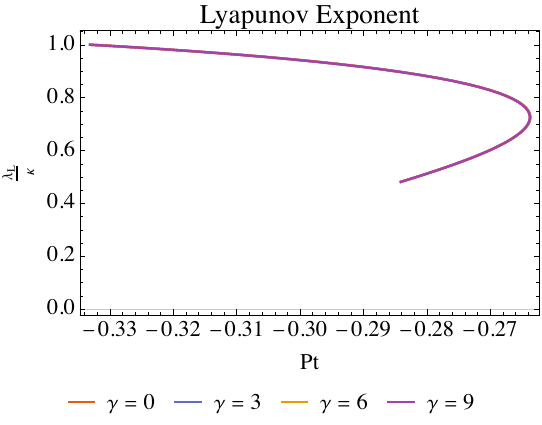}
    \includegraphics[width=0.45\linewidth]{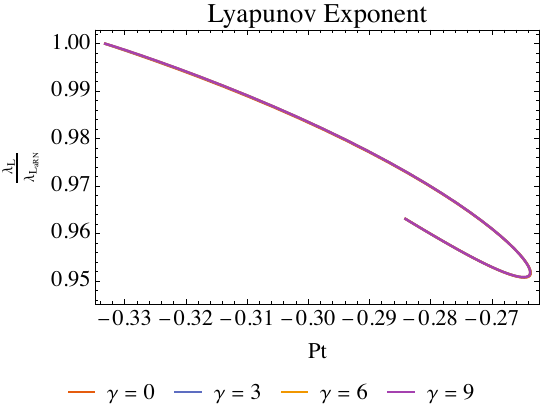}
    \caption{Plots of $\lambda_L/\kappa$ vs $p_t$ (left) and $\lambda_L/\lambda_{L_\text{aRN}}$ vs $p_t$ (right) by varying the EMS parameter $\gamma$. In this plot, we set $\zeta=0, \rho=0.2,q=0.1$.}
    \label{fig:LSLKKasnergamma}
\end{figure}

\begin{figure}
    \centering
    \includegraphics[width=0.45\linewidth]{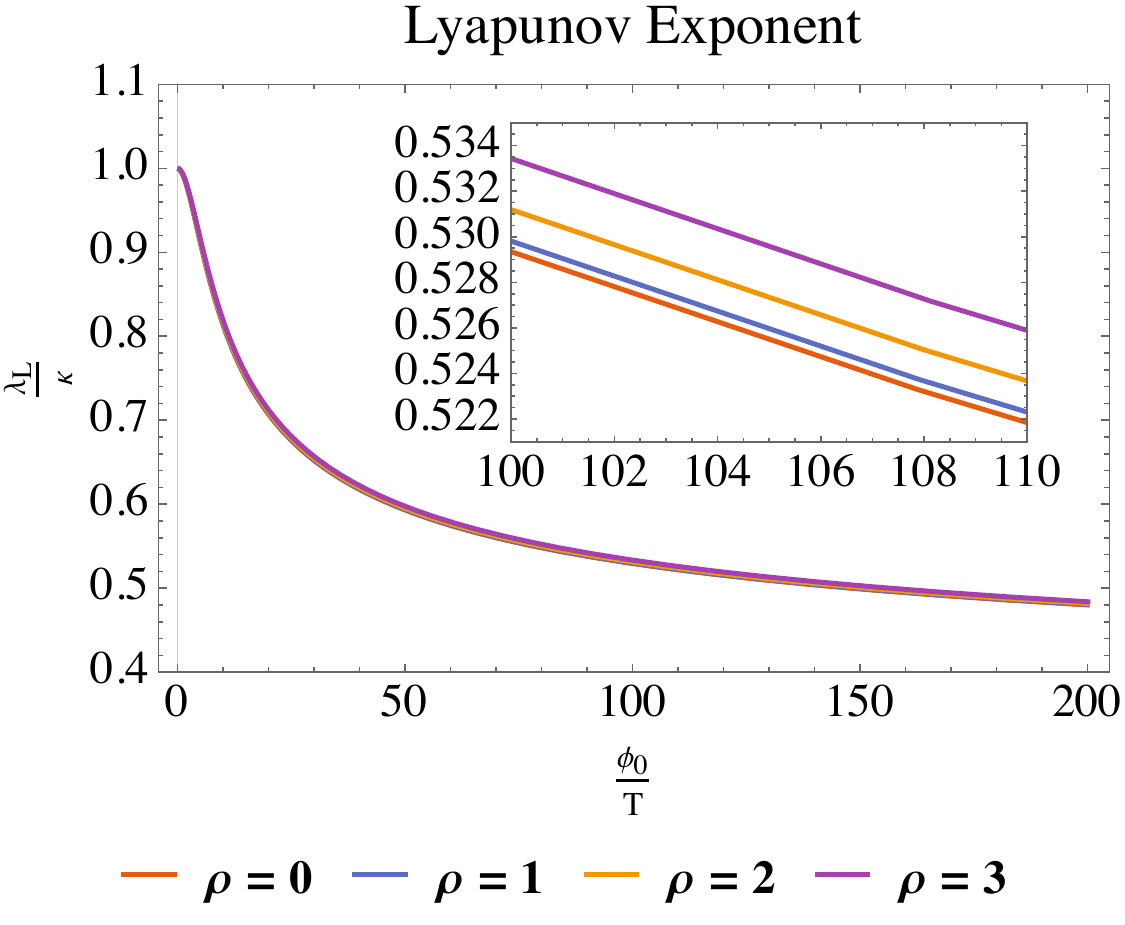}
    \includegraphics[width=0.45\linewidth]{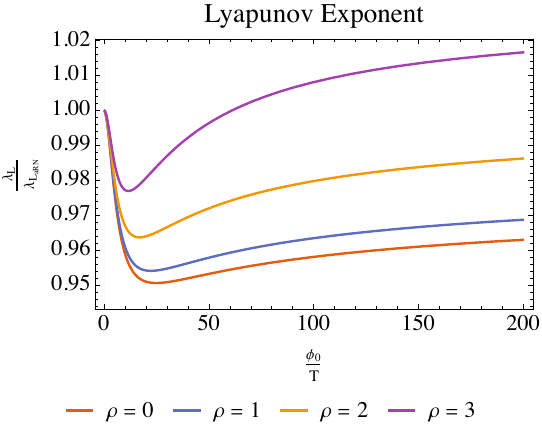}
    \caption{Plots of $\lambda_L/\kappa$ vs $\phi_0/T$ (left) and $\lambda_L/\lambda_{L_\text{aRN}}$ vs $\phi_0/T$ (right) by varying the charge density $\rho$. In this plot, we set $\zeta=0, \gamma=0 ,q=0.1$.}
    \label{fig:LSLKrho}
\end{figure}

\begin{figure}
    \centering
    \includegraphics[width=0.45\linewidth]{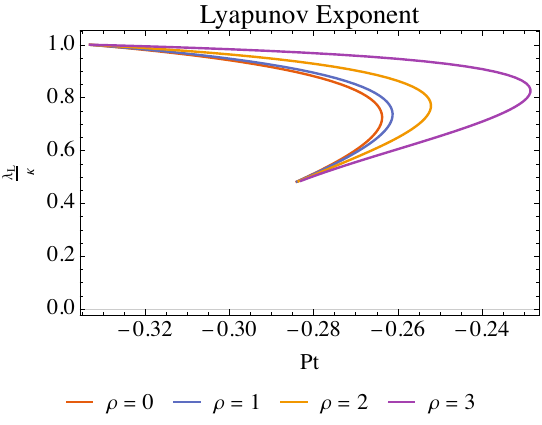}
    \includegraphics[width=0.45\linewidth]{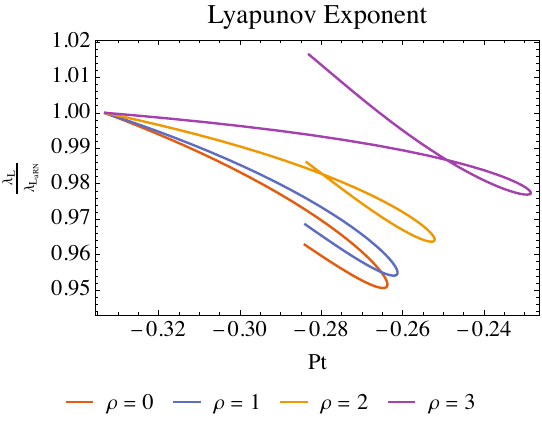}
    \caption{Plots of $\lambda_L/\kappa$ vs $p_t$ (left) and $\lambda_L/\lambda_{L_\text{aRN}}$ vs $p_t$ (right) by varying the charge density $\rho$. In this plot, we set $\gamma=0, \zeta=0 ,q=0.1$.}
    \label{fig:LSLKKasnerrho}
\end{figure}

The Lyapunov exponent that characterizes chaos in our system can be extracted from the area of the minimal surface $\mathcal{A_{A\cup B}}=4\mathcal{A}$. This area is proportional to the instantaneous Lyapunov exponent $\lambda_L$ at late times, where the insertion time is much larger than the thermal time, $t_w\gg\beta$ \cite{Malvimat2023KerrAdS4}. In this limit, the first term of $\mathcal{A}$ dominates and the area is proportional to the Lyapunov exponent as
\begin{equation}
    \mathcal{A}_{A\cup B}=\mathcal{A}_{A\cup B}^{(0)}\lambda_L t_w,
\end{equation}
where $\mathcal{A}_{A\cup B}^{(0)}$ is the unperturbed area. In some black holes with spherical horizons, the unperturbed area can be taken to be proportional to the black hole's entropy, since for large enough subsystems, the entanglement entropy scales as the entropy of the system. In our case, however, the black hole has a planar horizon and therefore the area is infinite.\\
\indent The area $\mathcal{A}_{A\cup B}$ is also proportional to the infinite length scale $L_y$ that needs to be canceled in calculating the Lyapunov exponent $\lambda_L$. Therefore, the unperturbed area $\mathcal{A}_{A\cup B}^{(0)}$ must also be linearly dependent on $L_y$ as well. Determining the exact value of the unperturbed area is rather difficult. However, we overcome this problem by imposing that in the undeformed case with $\phi_0\rightarrow 0$, the Lyapunov exponent reduces to the aRN case in the AdS background, which is nothing but the surface gravity of the black hole that saturates the Maldacena-Shenker-Stanford chaos bound \cite{Maldacena2016bound} for static black holes. Thus, we take the Lyapunov exponent to be
\begin{equation}
    \lambda_L=\frac{2}{\mathcal{N}}\qty[\frac{-f(r_c)e^{-\chi(r_c)}}{r_c^4}]^{1/2},
\end{equation}
where $\mathcal{N}$ is the normalization parameter that satisfy
\begin{equation}
 \mathcal{N}=\frac{2}{\kappa}\qty[\frac{-f(r_c)e^{-\chi(r_c)}}{r_c^4}]^{1/2}\bigg|_{\phi_0\rightarrow 0}.
\end{equation}
In this case, 
\begin{equation}
    \lambda_L(\phi_0\rightarrow0)=\kappa,
\end{equation}
as intended. We also absorb the infinite length $L_y$ into the definition of the normalization parameter.\\
\indent From this identification, we can plot the value of the Lyapunov exponent $\lambda_L$ as a function of the deformation parameter $\phi_0/T$. We mainly focus on the behavior of the Lyapunov exponent as a function of $\phi_0$, i.e., whether it increases or decreases compared to both $\kappa$ and the aRN value $\lambda_{L_\text{aRN}}$ and not on the actual value of the Lyapunov exponent. Therefore, we plot the ratio between the Lyapunov exponent and the surface gravity and also the ratio between $\lambda_L$ and $\lambda_{L_\text{aRN}}$ versus the boundary deformation parameter $\phi_0/T$. The aRN Lyapunov exponent can be taken to be
\begin{align}
    \lambda_{L_{\text{aRN}}}=\frac{2}{\mathcal{N}_{\text{aRN}}}\bigg[\frac{1}{r_c^4}\bigg(1&-\frac{r_c^3}{r_h^3}-\frac{r_c^2\zeta^2}{2}+\frac{r_c^3\zeta^2}{2r_h}+\frac{r_c^4\rho^2}{4}\\\nonumber
    &-\frac{r_c^3r_h\rho^2}{4}\bigg)\bigg]^{1/2},
\end{align}
where $\mathcal{N}_\text{aRN}$ is the aRN version of the normalization parameter to ensure $\lambda_{L}(\phi_0\rightarrow 0)=\lambda_{L_\text{aRN}}$. In the aRN case, the critical radius $r_c$ satisfy
\begin{equation}
    -\frac{4}{r_c^5}+\frac{\zeta ^2}{r_c^3}+\frac{\rho ^2 r_h^4-2 \zeta ^2 r_h^2+4}{4 r_c^2 r_h^3}=0.
\end{equation}
\begin{figure}
    \centering
    \includegraphics[width=0.8\linewidth]{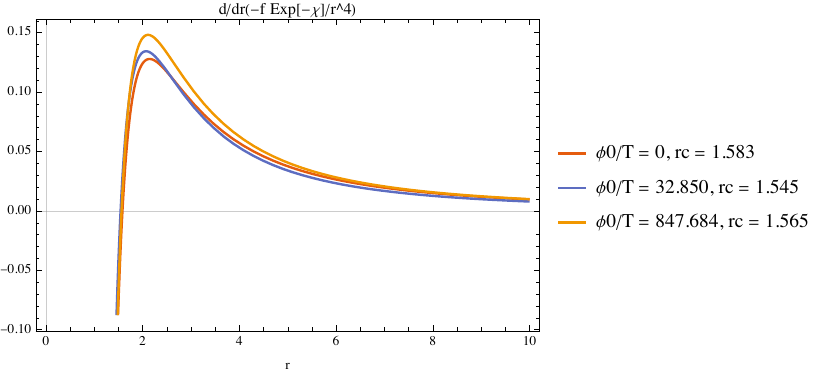}
    \caption{Plot for $\frac{d}{dr}\qty(\frac{-fe^{-\chi}}{r^4})$ versus $r$, where we use $\rho=2$, $\gamma=0.2$, $\zeta=0.3$, and $n=1$. The root of this plot corresponds to the critical radius $r_c$. As one can see, the root lies not far from the horizon $r_h=1$.}
    \label{fig:findroot}
\end{figure}
\indent In finding the critical radius $r_c$ for calculating the Lyapunov exponent $\lambda_L$, we use the function FindRoot in Mathematica to find the root around the horizon. As one can see in Fig. \ref{fig:findroot}, we can expect to not find another root in the deep interior since the graph asymptotes to zero in large $r$. However, one might anticipate that $r_c\rightarrow\infty$ also corresponds to
\begin{equation}
    \frac{d}{dr}\qty(\frac{fe^{-\chi}}{r^4})\bigg|_{r=r_c}\rightarrow0,
\end{equation}
and hence, taking the limit $r_t$ to be very close to the singularity will also render $\alpha\rightarrow\infty$. We will get back to this limit later, but for now, we focus on calculating $\lambda_L$ for $r_c$ near the horizon.\\
\begin{figure}
    \centering
    \includegraphics[width=0.45\linewidth]{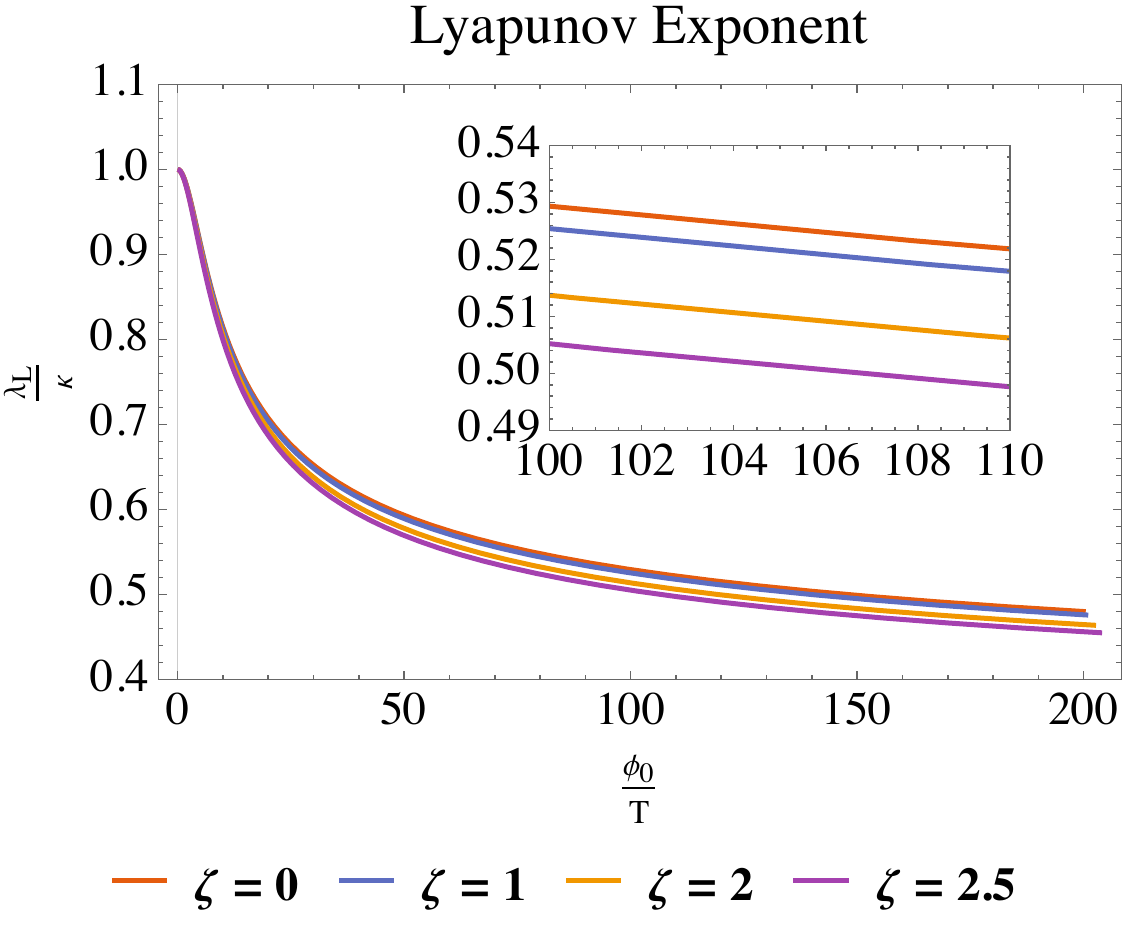}
    \includegraphics[width=0.45\linewidth]{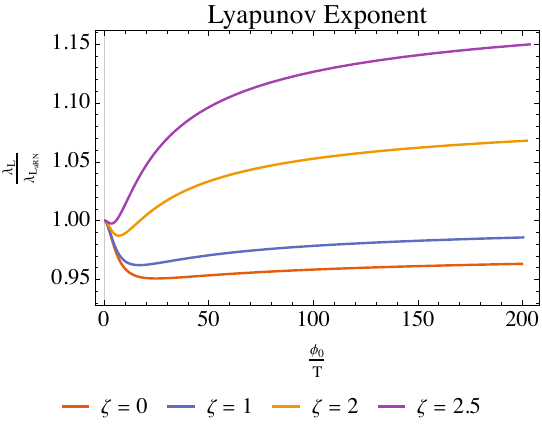}
    \caption{Plots of $\lambda_L/\kappa$ vs $\phi_0/T$ (left) and $\lambda_L/\lambda_{L_\text{aRN}}$ vs $\phi_0/T$ (right) for various value of the axion field parameter $\zeta$. In this plot, we set $\gamma=0, \rho=0.2,q=0.1$.}
    \label{fig:LKLSDeformationZeta}
\end{figure}
\indent We plot the ratio $\lambda_L/\kappa$ and $\lambda_L/\lambda_{L_{\text{aRN}}}$ with respect to the boundary deformation parameter $\phi_0/T$ to see how boundary deformation affects the Lyapunov exponent of the black hole. The ratio $\lambda_L/\kappa$ decreases monotonically as the deformation parameter $\phi_0/T$ increases, as can be seen in Figure \ref{fig:LKLSDeformationZeta} (left). We can justify this plot by noticing that $T\rightarrow0$ in the limit $\phi_0/T\rightarrow\infty$. This limit corresponds to an extremal black hole with zero temperature which also has a vanishing Lyapunov exponent (see \cite{Malvimat2023KerrAdS4,Prihadi2023}). This also means that the Lyapunov exponent $\lambda_L$ approaches zero faster than the surface gravity $\kappa$ in the limit $T\rightarrow0$. We cannot definitively conclude that $\lambda_L / \kappa$ approaches zero in the large $\phi_0 / T$ limit, as our numerical calculations are finite. However, we observe from our plot that it becomes smaller as $\phi_0 / T$ increases. In this plot, the axion parameter makes the vanishing property even faster as $T\rightarrow 0$.\\
\indent On the other hand, the ratio $\lambda_L/\lambda_{L_\text{aRN}}$ does not have monotonic behavior, as can be seen in Figure \ref{fig:LKLSDeformationZeta} (right). It decreases at first and starts to increase after some value of $\phi_0/T$. At some point, when $\zeta=2$ and $\zeta=2.5$ for example, the ratio is greater than one, meaning that a charged hairy black hole can be more chaotic than the aRN black hole. We also anticipate that as the deformation grows, the charged hairy black hole will eventually become more chaotic than the aRN black hole. Moreover, the axion parameter $\zeta$ appears to play a role in enhancing the chaotic behavior. The physics underlying this non-monotonic behavior remains unclear, as our results are based on numerical calculations. An analytical investigation, particularly through solving the equations of motion, is needed to provide a more detailed analysis of this behavior.\\
\indent We also study the dependence of the Lyapunov exponent $\lambda_L$ on the Kasner exponent $p_t$ by varying $\phi_0/T$. The plot of $\lambda_L/\kappa$ and $\lambda_L/\lambda_{L_\text{aRN}}$ versus $p_t$ can be seen in Figure \ref{fig:LSLKKasnerZeta}. Both figures show a non-invertible relation between $\lambda_L$ and $p_t$. This is already anticipated as the relation between $p_t$ and $\phi_0/T$ as shown in Figure \ref{fig:plotkasner} is also non-invertible. The behavior of the scrambling time $t_*$ and the butterfly velocity $v_B$ for a neutral hairy black hole also exhibits similar results \cite{caceres_shock_2024}. The non-invertible relation indicates that we cannot fully describe the interior geometry from boundary data alone and the Lyapunov exponent is no exception. Other than that, we are interested in how the parameters $\zeta,\gamma$, and $\rho$ affect the relation between $\lambda_L$ and $p_t$.\\
\indent We also do similar plots but varying the EMS parameter $\gamma$ (see Figures \ref{fig:LSLKgamma} and \ref{fig:LSLKKasnergamma}) and the charge density $\rho$ (see Figures \ref{fig:LSLKrho} and \ref{fig:LSLKKasnerrho}). We can see that the parameter $\gamma$, in this case, does not affect the plots while the charge density $\rho$ increases both $\lambda_L/\kappa$ and $\lambda_L/\lambda_{L_\text{aRN}}$. Although our result seems constant with respect to $\gamma$, the EMS parameter plays important role in shifting the scrambling time delay. This will be discussed in Section 4.\\
\indent Our results so far indicates that the Lyapunov exponent, being some quantities from the boundary theory, cannot completely determine Kasner geometry in the interior of the black hole since the relation is not invertible. This is so even though in calculating the Lyapunov exponent, we involve the calculation of the mutual information $I(A;B)$ that comes from the entangling Hartman-Maldacena surface that slightly penetrates the interior at $r=r_c$. Other quantity such as the subleading term near the singularity is needed to reconstruct interior geometry in terms of boundary data completely. Nevertheless, we obtain some important information on how the parameters $\zeta,\gamma,\rho$ affect the ratios of the Lyapunov exponent.
\subsection{Near-Singularity Area Functional}
To learn more about physics near the singularity in terms of the mutual information $I(A;B)$, we also calculate the area functional $\mathcal{A}$ defined in Eq. \eqref{areafuncitonal}, as $r_t\rightarrow\infty$. In this limit, the conserved quantity $K$ approaches zero, as well as the boundary time $t_b$ from Eq. \eqref{timevsr}. \\
\indent From Eq. \eqref{Kconstant}, and the near-singularity expansions of $f$ and $\chi$, we obtain the relation between the turning point $r_t$ and $K$, as $r_t\rightarrow\infty$,
\begin{equation}
    r_t=\qty(\frac{f_{K1}e^{-\chi_{K1}}}{K^2})^{\frac{1}{1+c^2}}+...\;,
\end{equation}
where $...$ denotes subleading terms as $K\rightarrow0$. Using this relation, the area functional then becomes
\begin{equation}
    \mathcal{A}=L_y\int dr\frac{1}{\sqrt{-f_{K1}}r^{(7+2c^2)/2}}\qty[1-\qty(\frac{K^2}{f_{K1}e^{-\chi_{K1}}})r^{1+c^2}]^{1/2},
\end{equation}
in one of its end point near the singularity. This integral can be calculated explicitly using binomial expansion and it gives us
\begin{equation}
    \mathcal{A}=\tilde{l}K^{\frac{2c^2+5}{c^2+1}},
\end{equation}
where $\tilde{l}$ is a constant that depends on $L_y$ and other near-singularity data such as $f_{K1},\chi_{K1},c$ and does not depends on $K$. This term of the area functional approaches zero as $K\rightarrow0$.\\
\indent Aside from this endpoint, we also need to calculate the area functional in the other endpoint which is closer to the boundary. Although it depends on boundary data such as $\phi_0,\expval{\mathcal{O}},\expval{T_{tt}}$, it only depends linearly on $K$ and thus also vanishes in $K\rightarrow0$ limit. The explicit calculation for both endpoints after we subtract the universal divergent part gives us
\begin{equation}
    \mathcal{A}=l K+...+\tilde{l}K^{\frac{2c^2+5}{c^2+1}}+...\;,
\end{equation}
where $l$ depends on the entire flow from the boundary to the singularity and $...$ denotes higher order of $K$ so that the area vanishes in both its endpoints. The vanishing area functional $\mathcal{A}$ indicates the time scale where the mutual information $I(A;B)$ is still large and thus the OTOC is also still large. In this time scale, we do not expect to observe any exponential decay behavior of the OTOC and thus it may not be suitable for extracting information about the chaotic properties of the system. However, since this area penetrates deep into the interior, we expect to see something interesting about its relation to the emergence of Kasner spacetime. Future investigations, including more rigorous analytical analyses, may shed further light on these aspects.
\subsection{Localized Chaos and Butterfly Velocity}
\begin{figure}
    \centering
    \includegraphics[width=0.32\linewidth]{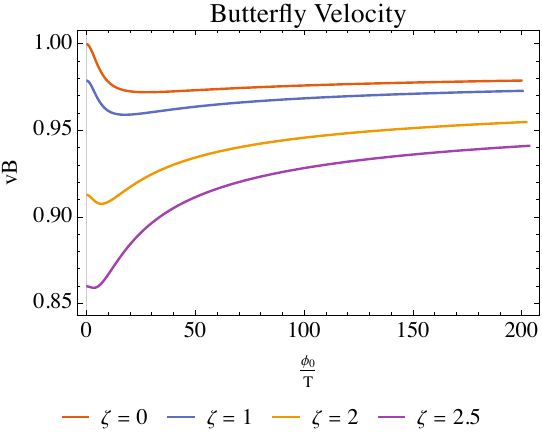}
    \includegraphics[width=0.32\linewidth]{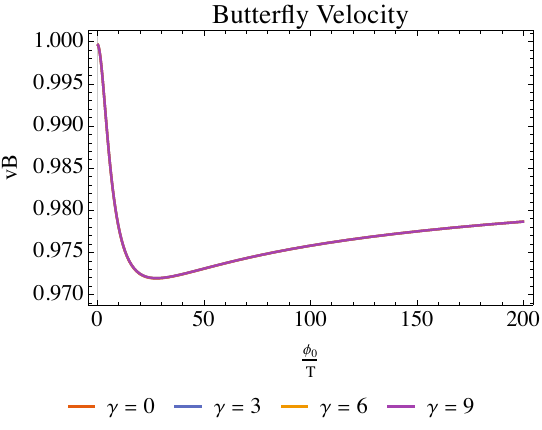}
    \includegraphics[width=0.32\linewidth]{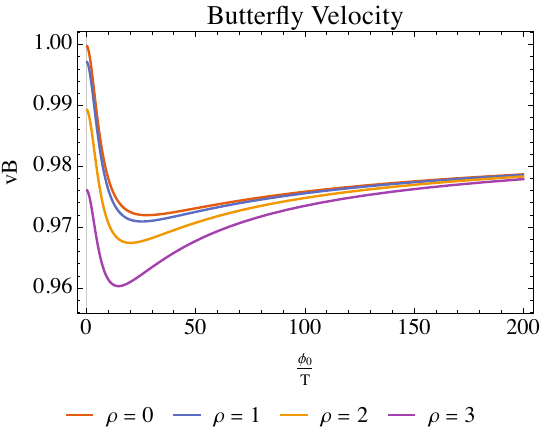}
    \caption{Plot of the butterfly velocity $v_B$ versus the boundary deformation $\phi_0/T$ for various values of $\zeta$ (left), $\gamma$ (center), and $\rho$ (right). The butterfly velocity is normalized with respect to its Schwarzschild value. In this plot, we use $q=0.1$}
    \label{fig:BVDeformation}
\end{figure}

\begin{figure}
    \centering
    \includegraphics[width=0.32\linewidth]{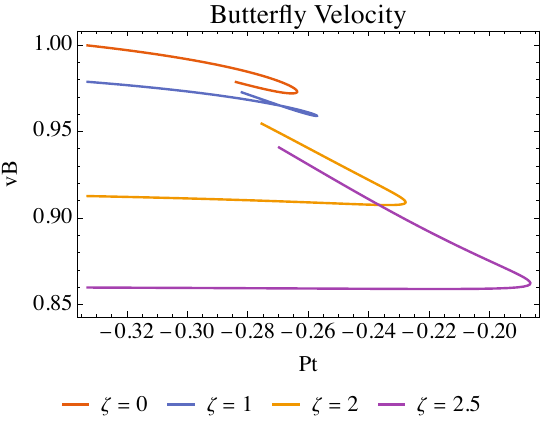}
    \includegraphics[width=0.32\linewidth]{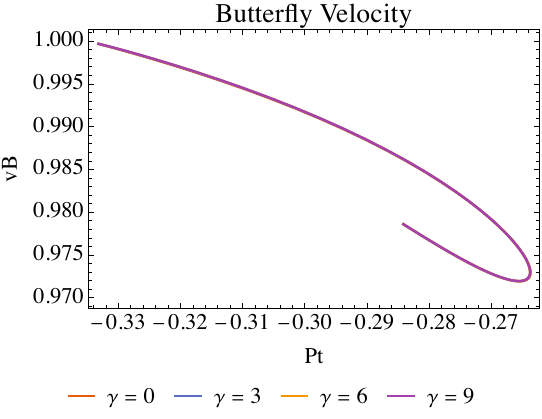}
    \includegraphics[width=0.32\linewidth]{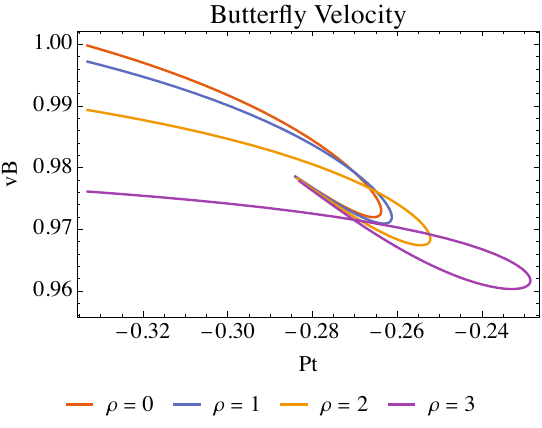}
    \caption{Plot of the butterfly velocity $v_B$ versus the Kasner exponent $p_t$ for various values of $\zeta$ (left), $\gamma$ (center), and $\rho$ (right). The butterfly velocity is normalized with respect to its Schwarzschild value.}
    \label{fig:BVKasner}
\end{figure}

In this section, we find out how the parameters $\zeta,\gamma,\rho$ affect the butterfly velocity as a function of boundary deformation $\phi_0/T$ and Kasner exponent $p_t$. If the gravitational shock waves are sent from the left boundary at some local point in space, the shock wave parameter $\alpha$ depends locally on position coordinate $x,y$ and propagates in a null path at $U=0$. For simplicity, we may assume that the gravitational shock waves function is in the form of $\delta(\vec{x})$. For large insertion time $t_w$, the gravitational shock waves are boosted as it approaches the black hole horizon, and the stress-energy tensor is given by \cite{Prihadi2024}
\begin{equation}
    T_{UU}^{\text{shock}}=Ee^{\frac{2\pi t_w}{\beta}}\delta(U)\delta(\vec{x}).
\end{equation}
The gravitational shock waves perturb the background geometry as $dV\rightarrow dV+\delta(U)\alpha(x)dU$, giving us the Dray-'t Hooft solution. Plugging in this solution to the Einstein's equation with $T_{UU}^{\text{shocks}}$ gives us (see \cite{caceres_shock_2024} for more detailed expressions)
\begin{equation}
    \alpha(x)=\frac{e^{\frac{2\pi}{\beta}\qty((t_w-t_*)-\frac{|x|}{v_B})}}{|x|^{1/2}},
\end{equation}
where
\begin{equation}\label{butterflyvelocity}
    v_B=\sqrt{\pi T r_he^{-(\chi(r_h)-\chi(0))/2}},
\end{equation}
is defined as the butterfly velocity.\\
\indent Again, this butterfly velocity is influenced by the boundary deformation especially through the function $\chi$. The Schwarzschild value of the butterfly velocity is $v_B=\sqrt{3}/2$. We normalize our calculations with respect to the Schwarzschild value so that it reduces to one when both the axion paramater $\zeta$ and the boundary deformation $\phi_0$ vanish. The dependence on the butterfly velocity with the boundary deformation parameter $\phi_0/T$ can be seen in Figure \ref{fig:BVDeformation}. We see that both $\zeta$ and $\rho$ decreases the value of $v_B$. At $\phi_0\rightarrow0$, the butterfly velocity shows us its aRN value rather than the Schwarzschild value $v_B=\sqrt{3}/2$. Furthermore, the relation between $v_B$ and the Kasner exponent $p_t$ can be seen in Figure \ref{fig:BVKasner}.
\section{Shock Wave Bounce in the Interior and Scrambling Time Delay}
\begin{figure}
    \centering
    \includegraphics[width=0.32\linewidth]{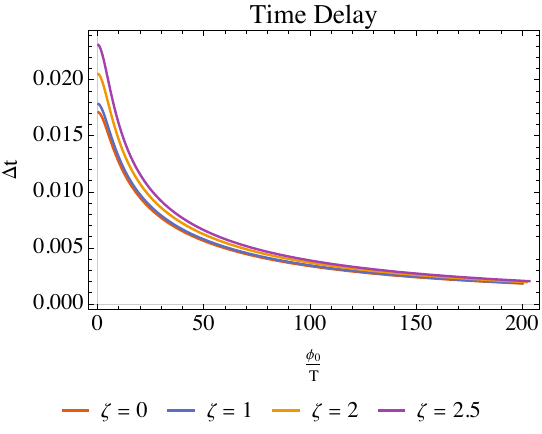}
    \includegraphics[width=0.32\linewidth]{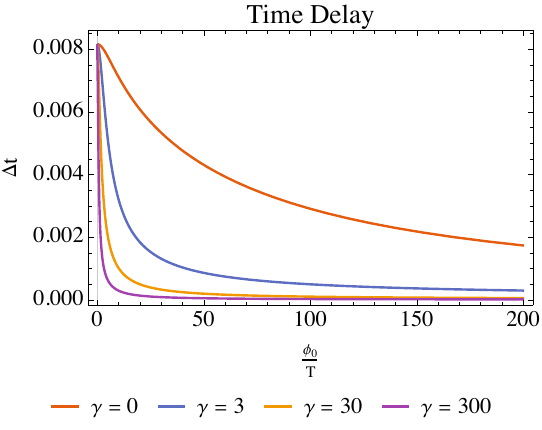}
    \includegraphics[width=0.32\linewidth]{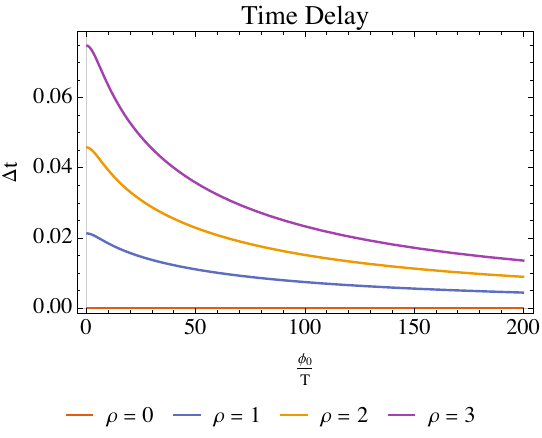}
    \caption{Plot of the scrambling time delay $\Delta t$ versus the boundary deformation $\phi_0/T$ for various values of $\zeta$ (left), $\gamma$ (center), and $\rho$ (right). In this plot, we use $q=0.1$ and $\mathcal{Q}=1$.}
    \label{fig:Delay}
\end{figure}

\begin{figure}
    \centering
    \includegraphics[width=0.45\linewidth]{DelayGamma1.pdf}\includegraphics[width=0.45\linewidth]{DelayRho1.pdf}\\
    \includegraphics[width=0.45\linewidth]{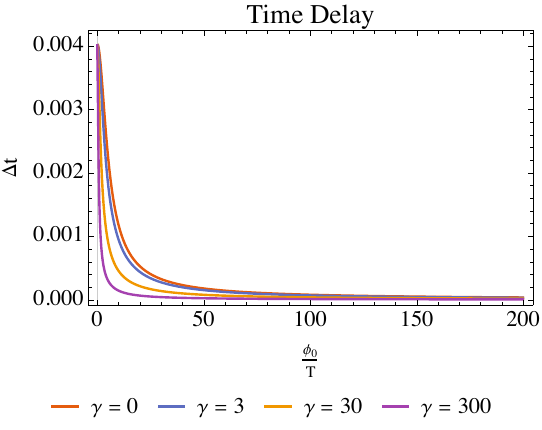}\includegraphics[width=0.45\linewidth]{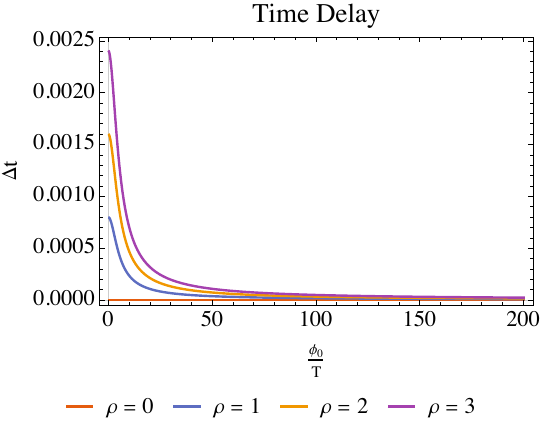}\\
    \includegraphics[width=0.45\linewidth]{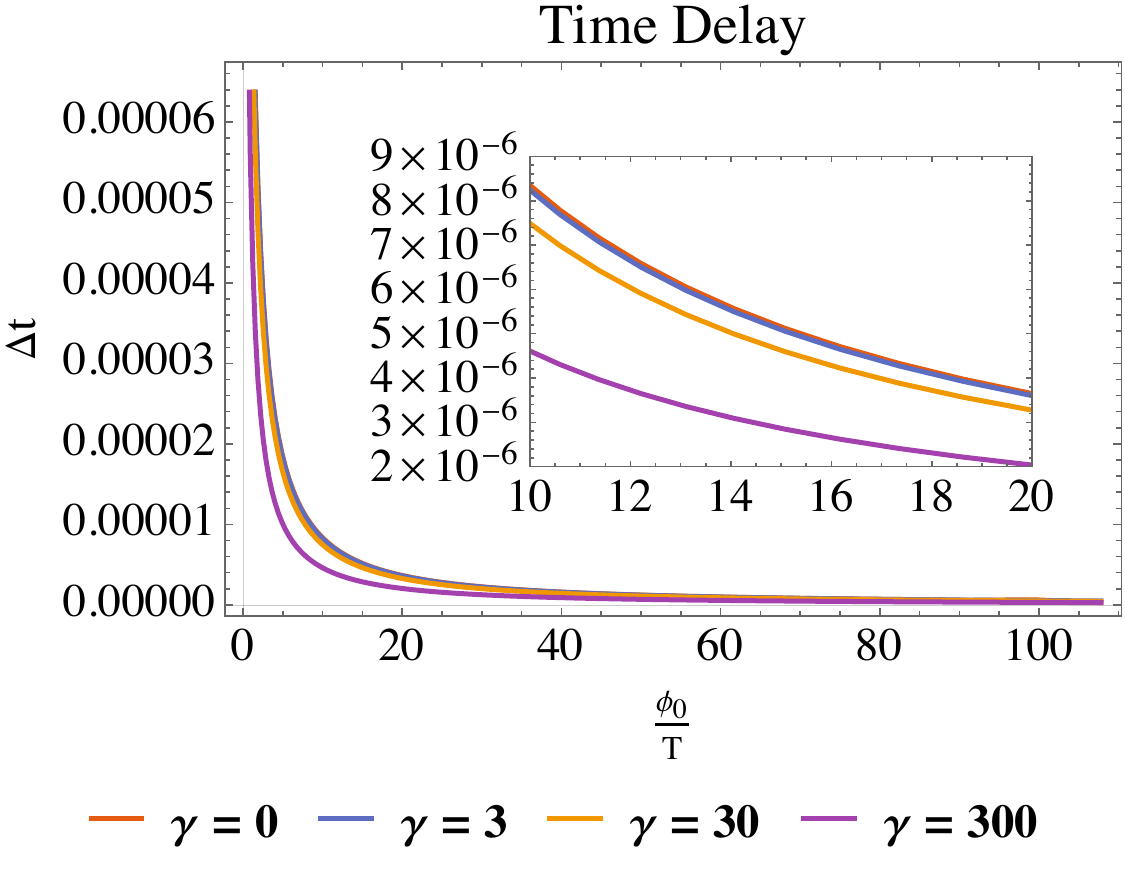}\includegraphics[width=0.45\linewidth]{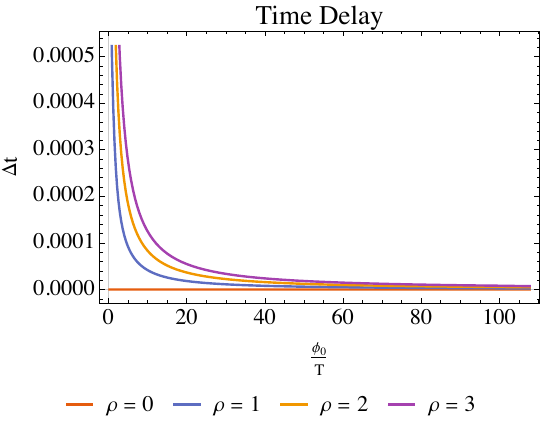}
    \caption{Plot of $\Delta t$ vs $\phi_0/T$ with $q=0.1$ (top), $q=1$ (middle), and $q=5$ (bottom). We vary the values of $\gamma$ (left) and $\rho$ (right). In this plot, we use $\mathcal{Q}=0.5$.}
    \label{fig:Delayq}
\end{figure}

Since we perturb our black hole by charged shock waves instead of neutral ones, we may expect bouncing in the interior as first explained in \cite{Horowitz2022}. This bounce happens so that the null energy condition is not violated. We are particularly interested in investigating this bounce that leads to scrambling time delay and its relation to the boundary deformation and the Kasner exponents, as this bounce also occurs in the interior, albeit near the horizon.\\
\indent The mutual information $I(A;B)$ in Eq. \eqref{mutualinformation} vanishes when the insertion time $t_w$ is given by
\begin{equation}
    \kappa t_*=\log S+\frac{(\mathcal{A}_A+\mathcal{A}_B)}{4\pi\sqrt{-f(r_c)e^{-\chi(r_c)}/r_c^4}}+\log\frac{1}{1-\mu\mathcal{Q}}.
\end{equation}
This time scale $t_*$ is defined as the scrambling time. The first term indicates that the system is fast scrambling, while the second term can be neglected in the large $S$ limit since it does not scale as the entropy of the black hole. The last term comes from the interaction between the shock wave charge per unit of energy $\mathcal{Q}$ and the black hole's electric potential. In other works such as \cite{Prihadi2023}, the black hole's electric potential is given by the value of $\Phi(r)$ evaluated at the horizon, $\Phi(r_h)$. In our case, we set $\Phi(r_h)=0$ so that the chemical potential $\mu$ is given by the boundary value of $\Phi(r)$. Recent work \cite{Horowitz2022} shows that the last term corresponds to the delay of the scrambling process, instead of prolonging the scrambling time.\\
\indent When either the shock wave's or black hole's charge vanishes, this term also vanishes. We can see how the boundary deformation $\phi_0$ affects the scrambling time delay in Figure \ref{fig:Delay}. In general, $\phi_0/T$ decreases the scrambling time delay. From the plots, we can see that the axion parameter $\zeta$ and $\rho$ increase the scrambling time delay while the EMS parameter $\gamma$ decreases it. It is in this plot that the EMS coupling $\gamma$ plays an important role. This is because $\gamma$ corresponds to a direct coupling between the scalar field $\phi^2$ and the Maxwell field through $F_{\mu\nu}F^{\mu\nu}$ in the action. The charge density $\rho$ corresponds to the black hole's charge and hence it plays the central role in creating the shock wave bounce. When $\rho=0$, the black hole is uncharged. Therefore, even though the gravitational shock wave is charged, it does not interact with anything in the interior and there is no time delay. Although it is not directly coupled to the electric potential $\Phi$, the axion parameter $\zeta$ also affects the scrambling time delay. For a larger value of $\zeta$, the scrambling time delay is increased. \\
\indent When the EMS coupling $\gamma$ is large, we observe that the scrambling time delay rapidly approaches zero as the boundary deformation increases. This suggests that strong EMS coupling accelerates the scrambling process, leading to an effectively instantaneous start of the scrambling process. A possible explanation for this behavior lies in the equation of motion containing the coupling between $\phi^2$ and the electric potential $\Phi^2$. Since this term directly affects how the boundary deformation determines the chemical potential $\mu$ in the boundary theory, increasing the EMS coupling may cause $\mu$ to change more significantly with the deformation. This, in turn, could influence the near-horizon dynamics in such a way that the scrambling delay is suppressed. \\
\indent Furthermore, this also explains why the EMS parameter primarily affects the scrambling time delay but not other chaotic properties such as the quantum Lyapunov exponent or butterfly velocity. Since the scrambling time delay is directly related to \( \mu \), any modification in the EMS coupling alters the delay through its impact on the boundary chemical potential. In contrast, quantities like the Lyapunov exponent and butterfly velocity are determined by more universal properties of the near-horizon dynamics, which are not explicitly dependent on \( \mu \). This distinction highlights the unique role of the EMS coupling in controlling scrambling time.\\
\indent Other than the parameters $\zeta,\gamma,\rho$, the scrambling time delay should also be influenced by the coupling $q$. In figure \ref{fig:Delayq}, we plot the time delay for $q=\{0.1,1,5\}$. The result shows that for larger $q$, the curves for different values of $\gamma$ and $\rho$ become tighter, which agree with the results in \cite{sword_kasner_2022}. This might be related to the probe limit $q\gg1$, where the backreaction of the probe on the background geometry becomes negligible \cite{hartnoll_building}, leading to a universal scrambling behavior that is less sensitive to variations in $\gamma$ and $\rho$. In this regime, the dynamics of the perturbation are primarily governed by the background structure rather than the details of the probe, which could explain why the curves become tighter for larger $q$.
\begin{figure}
    \centering
    \includegraphics[width=0.32\linewidth]{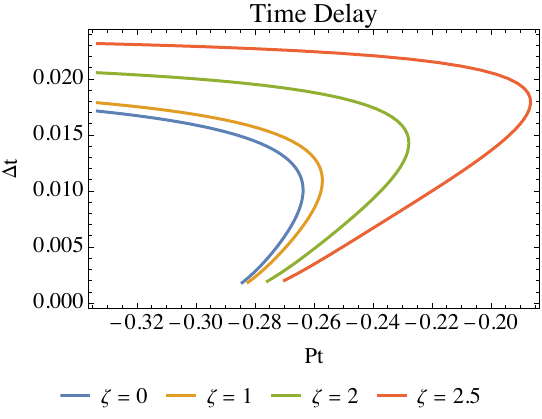}
    \includegraphics[width=0.32\linewidth]{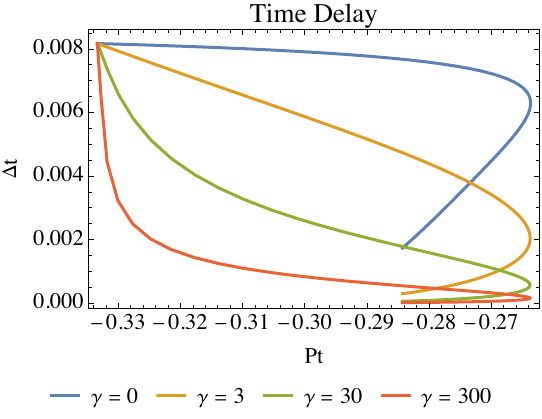}
    \includegraphics[width=0.32\linewidth]{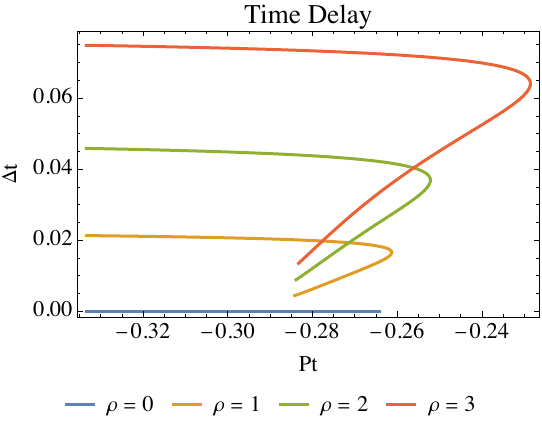}
    \caption{Plot of the scrambling time delay $\Delta t$ versus the Kasner exponent $p_t$ for various values of $\zeta$ (left), $\gamma$ (center), and $\rho$ (right). In this plot, we use $q=0.1$ and $\mathcal{Q}=1$.}
    \label{fig:DelayKasner}
\end{figure}
\section{Summary and Discussions}
In this work, we study how chaotic parameters—such as the quantum Lyapunov exponent $\lambda_L$, the butterfly velocity $v_B$, and the scrambling time delay $\Delta t$—depend on the dimensionless deformation parameter $\phi_0/T$, which induces a more general Kasner geometry in the interior of a charged hairy black hole. We obtain these chaotic parameters by injecting charged gravitational shock waves into the black hole, which undergo exponential blueshifting as they cross the horizon. The boundary deformation affects the Kasner geometry in the interior as characterized by the Kasner exponents. Our goal is to understand how boundary parameters provide insight into the interior Kasner geometry through their relation to the boundary deformation paramter $\phi_0/T$ and the Kasner exponents.\\
\indent We consider a general charged hairy black hole with an axion parameter $\zeta$ and an Einstein-Maxwell-Scalar (EMS) coupling $\gamma$, which introduces an interaction term of the form $F_{\mu\nu}F^{\mu\nu}|\phi|^2$. The bulk scalar field $\phi(r)$ also interacts with the bulk Maxwell field $A_\mu$, characterized by the coupling $q$ and charge density $\rho$. Our analysis shows that the ratio of the quantum Lyapunov exponent to the surface gravity decreases as the deformation parameter $\phi_0$ increases, though for sufficiently large deformation, the Lyapunov exponent in the deformed geometry can exceed that of the axion Reissner-Nordström case. Moreover, the butterfly velocity exhibits a nontrivial dependence on the deformation. \\
\indent The scrambling time $t_*$ receives a contribution that depends on the interaction between the chemical potential $\mu$ and the shock wave charge per unit energy $\mathcal{Q}$. This contribution, first studied in \cite{Horowitz2022}, corresponds to the scrambling time delay. We analyze how the parameters $\zeta, \gamma, \rho$ affect this time delay and its relation to the boundary deformation parameter $\phi_0/T$ and the Kasner exponent $p_t$. Our results show that the scrambling time delay $\Delta t$ decreases as the boundary deformation increases, with this effect being particularly sensitive to the EMS coupling $\gamma$. Notably, for large values of $\gamma$, the scrambling time delay can become very small once the boundary deformation is turned on, suggesting that the EMS coupling plays a significant role in determining how fast the black hole scrambling process begins even though both the black hole and the gravitational shock wave are charged. Although it does not directly control the coupling between $\phi$ and $\Phi$, the axion parameter $\zeta$ slightly affects $\Delta t$. As $\zeta$ increases, the time delay also becomes longer. This behavior is similar to the effect of increasing $\rho$, as a larger charge density also significantly prolongs the scrambling time delay. This is expected since $\rho$ is the main factor responsible for the bounce in the black hole interior with greater $\rho$ corresponds to stronger interactions between the charged shock waves and the charged black hole.\\
\indent It has been shown \cite{sword_kasner_2022,Hartnoll2020,Cai2021} that the charged hairy black hole does not have an inner horizon. On the other hand, it is also known that the charged black hole solution (with $\phi=0$) has inner horizon instability due to mass inflation (see, for example, \cite{PhysRevLett.63.1663,Dokuchaev2014,ghosh2024} and references therein). It is worth noting that the absence of an inner horizon in the charged hairy black hole geometry might be related to the suppression of the mass inflation effect. Although our numerical results for chaos-related quantities such as the Lyapunov exponent smoothly approach those of the axion-Reissner-Nordström solution in the small-$\phi_0$ limit, the connection between this limit and the suppression of mass inflation near the inner horizon remains unclear. We believe that future work in this direction could shed important light on the problem. Several earlier studies (for example, [21,39]) have analytically shown that solutions with $f(r_I)=0$, where $r_I$ denotes the would-be inner horizon, do not arise in charged hairy black holes. These works could serve as a starting point for investigating the relation between the absence of the inner horizon and the suppression of mass inflation.\\
\indent In this work, we restrict our analysis to $L=1$ case for numerical simplicity, where $L$ is the AdS radius. However, learning about the large-$L$ limit is also interesting since it might coincide with the asymptotically flat black holes. In our previous work \cite{Prihadi2023}, we see that when the AdS radius becomes large, the minimal instantaneous Lyapunov exponent approaches its surface gravity, thus saturating the Maldacena-Shenker-Stanford bound. In our current setup, the numerical calculations fail when the AdS radius is dramatically increased. However, we may find interesting results if we could study the solution of the hairy black hole under the large-$L$ expansion and compare the result to chaos in flat-space holography \cite{Afshar2020}. Asymptotically flat black holes are of particular interest due to their relevance to astrophysical observations. Although still speculative, certain chaotic properties of black holes such as the Lyapunov exponent derived from particle orbits may have connections to observable features like black hole imaging \cite{Tavlayan2025}. Therefore, investigating the relation between the flat-space limit of holographic calculations and the chaotic behavior of particle motion around black holes could provide valuable insights for future studies.\\
\indent Another interesting future direction is to study how rotating shock waves \cite{Malvimat2022}, or even rotating and charged shock waves \cite{Prihadi2023}, disrupt the interior structure of rotating and charged hairy black holes \cite{Gao2024}. Understanding how these perturbations affect the black hole interior could provide new insights into the role of rotation and charge in chaotic dynamics. Furthermore, we are also interested in investigating how this relevant deformation influences the traversability of a wormhole induced by double-trace deformations on the boundary \cite{Gao2017,Ahn2024}. Additionally, recent work on analytical solutions of scalar-hairy black holes \cite{Atmaja2024} provides a foundation for further studies of chaotic behavior in specific limits, such as the low-temperature limit $T \to 0$. We hope this finding can help us in investigating the chaotic behavior especially in some certain limit of hairy black holes. We leave these investigations in future works.
\section*{Acknowledgement}
H. L. P. would like to thank Geoffry Gifari for early collaboration in numerical calculations. This work was done in part during the workshop "Holographic Duality and Models of Quantum Computation" held at Tsinghua Southeast Asia Center in Bali, Indonesia (2024). H. L. P. and F. K. would like to thank Veronika Hubeny, Sumit Das, Alexander Jahn, Charles Cao for helpful discussions during this workshop. H. L. P. would like to thank the National Research and Innovation Agency (BRIN) for its financial support through the Postdoctoral Program. F. K. would like to thank the Ministry of Education and Culture (Kemendikbud) Republic of Indonesia for financial support through Beasiswa Unggulan. D. D. is supported by the APCTP (YST program) through the Science and Technology Promotion Fund and Lottery Fund of the Korean Government and the Korean Local governments, Gyeongsangbuk-do Province, and Pohang city. F. P. Z. would like to thank riset PPMI, Fakultas Matematika dan Ilmu Pengetahuan Alam, Institut Teknologi Bandung and the Ministry of Higher Education, Science, and Technology (Kemendikti Saintek) for financial support.
\bibliographystyle{elsarticle-num}
\bibliography{BIBInterior.bib}

\end{document}